\documentclass[pra,amsmath,amssymb,amsfonts,nofootinbib]{revtex4}

\usepackage{bm,graphicx,mathrsfs}

\usepackage{amsmath,bbm}
\usepackage{amsfonts,amssymb}
\usepackage{graphicx}
\usepackage{epsfig}
\usepackage{times}
\usepackage{verbatim}

\newcommand{\be}{\begin{equation}}
\newcommand{\ee}{\end{equation}}
\newcommand{\bq}{\begin{eqnarray}}
\newcommand{\eq}{\end{eqnarray}}
\newcommand{\bea}{\begin{eqnarray}}
\newcommand{\eea}{\end{eqnarray}}
\newcommand{\ba}{\begin{align}}
\newcommand{\ea}{\end{align}}

\newcommand{\1}{\mathbbm{1}}

\newcommand{\ket}[1]{ | \, #1 \rangle}
\newcommand{\bra}[1]{ \langle #1 \,  |}
\newcommand{\oket}[1]{ | \, #1  ) }
\newcommand{\obra}[1]{ ( #1 \,  |}

\newcommand{\obraket}[2]{\left(\, #1\,|\,#2\,\right)}

\newcommand{\oproj}[1]{\oket{#1}\obra{#1}}

\newcommand{\avr}[1]{\left \langle#1 \right \rangle}

\newcommand{\tr}[1]{{\rm tr}\left[{#1}\right]}
\newcommand{\ptr}[2]{{\rm tr}_{#1}\left[{#2}\right]}

\newcommand{\raw}{\rightarrow}

\newcommand{\bR}{\mathbbm{R}}

\newcommand{\bC}{\mathbbm{C}}
\newcommand{\bZ}{\mathbbm{Z}}

\newcommand{\cH}{\mathcal{H}}
\newcommand{\cE}{\mathcal{E}}
\newcommand{\cV}{\mathcal{V}}

\newcommand{\cL}{\mathcal{L}}

\newcommand{\cS}{\mathcal{S}}

\newcommand{\cM}{\mathcal M}

\newcommand{\cO}{\mathcal O}
\newcommand{\cX}{\mathcal X}

\newcommand{\cP}{\mathcal P}
\newcommand{\cG}{\mathcal G}

\newcommand{\half}{\frac{1}{2}}

\newcommand{\Var}{{\rm Var}}

\newtheorem{theorem}{Theorem}
\newtheorem{lemma}[theorem]{Lemma}

\newtheorem{proposition}[theorem]{Proposition}
\newtheorem{definition}[theorem]{Definition}

\def\Proof{\noindent\textsc{Proof:}}
\def\proof{\Proof}
\def\qed{\leavevmode\unskip\penalty9999 \hbox{}\nobreak\hfill
    \quad\hbox{\leavevmode  \hbox to.77778em{%
               \hfil\vrule   \vbox to.675em%
               {\hrule width.6em\vfil\hrule}\vrule\hfil}}
     \par\vskip3pt}

\newcommand{\Sp}{\,\,\,\,\,\,}
\newcommand{\no}{\nonumber\\}

\begin{document}

\title{\sc \large Thermalization time bounds for Pauli stabilizer Hamiltonians}
\author{Kristan Temme  \\}
\address{Institute for Quantum Information and Matter, California Institute of Technology, Pasadena, CA 91125, USA}
\date{\today}

\begin{abstract}
We prove a general lower bound to the spectral gap of the Davies generator for Hamiltonians that can be written as the sum of commuting Pauli operators. These Hamiltonians, defined on the Hilbert space of $N$-qubits, serve as one of the most frequently considered candidates for a self-correcting quantum memory. A spectral gap bound on the Davies generator establishes an upper limit on the life time of such a quantum memory and can be used to estimate the time until the system relaxes to thermal equilibrium when brought into contact with a thermal heat bath. The bound can be shown to behave as $\lambda \geq \cO(N^{-1}\exp(-2\beta \, \overline{\epsilon}))$, where $\overline{\epsilon}$ is a generalization of the well known energy barrier for logical operators. Particularly in the low temperature regime we expect this bound to provide the correct asymptotic scaling of the gap with the system size up to a factor of $N^{-1}$. Furthermore, we discuss conditions and provide scenarios where this factor can be removed and a constant lower bound can be proven.
\end{abstract}

\maketitle

\section{Introduction}
A fundamental challenge in quantum information science is the protection of quantum information from decoherence. A proposed solution  \cite{shor1995scheme,dennis2002topological}  to this problem has been to encode the quantum information into a many-body entangled state and protect it this way from the action of local noise. This proposal lead to a new research field, referred to as quantum error correction \cite{gottesman1998theory,gottesman1997stabilizer}.  It has turned out that many ideas from quantum error correction have become increasingly useful in the theory of condensed matter physics \cite{yoshida2011classification}, as they help to understand new phases of quantum matter \cite{wen2004quantum}. One of the central questions in this field is that of thermal stability \cite{dennis2002topological,Zohar1,Zohar2,terhal,alicki2010thermal,alicki2009thermalization}.  Thermal stability plays a role in both the understanding of the behavior of topologically ordered systems at finite temperature, as well as in the estimation of the life time of self-correcting quantum memories. A standard  approach to self-correcting quantum memories is to encode the quantum information into the ground state, or any other suitable, subspace of a Hamiltonian. The Hamiltonian should have the property of shielding this subspace from thermal excitations.  An important, and also frequently studied, class of models are so-called stabilizer Hamiltonians. These Hamiltonians are directly related to stabilizer quantum codes \cite{gottesman1997stabilizer} and are given by the sum of commuting multi-qubit Pauli operators. 

In this paper, we will derive thermalization time bounds, also-called mixing time bounds, for the Davies generators \cite{Davis,Davis2}  of these Hamiltonians.  Davies generators are given in the form of a Lindblad equation \cite{Lindblad} and are known to converge to the Gibbs distribution of the particular Hamiltonian for which they are derived.   

The first rigorous upper bound on the memory time of a stabilizer Hamiltonian was derived for the two dimensional toric code model \cite{kitaev2003fault} in \cite{alicki2009thermalization}. The authors first proved a constant lower bound for the spectral gap of the Davies generator of the one dimensional Ising model. This bound could then be related to the spectral gap of the Davies generator of the toric code through a suitable partitioning of the two dimensional lattice. Other no-go results for stabilizer quantum memories \cite{bravyi2009no,landon2013local,haah2012logical} in lower dimensions rely on the absence of an energy barrier that separates two logical states in the code space. The argument proceeds to connect the energy barrier to the memory's life time through the phenomenological Arrhenius law $t_{mem}  \sim e^{\beta E_B}$, where $E_B$ is the energy barrier of the code \cite{bravyi2009no,brown2014quantum}. It has been an open question, whether there is in fact a rigorous connection between the energy barrier $E_B$ and thermalization time of the quantum system. Recent results \cite{bravyi2013quantum,michnicki20143d,yoshida2014violation} indicate that this law can only serve as an upper bound to the life time of the quantum memory.\\

The main result of this paper is a rigorous upper bound on the thermalization time of a qubit stabilizer Hamiltonian in terms of a quantity that can be seen as a generalization of the energy barrier $E_B$. The result is stated in theorem \ref{Gen-bound} in section \ref{sec:bound}. We estimate the thermalization time by finding lower bounds on the spectral gap of the Davies generators. The lower bounds on the spectral gap can be related to estimates of the trace-norm distance between any initial state and the thermal state of the stabilizer Hamiltonian. We show that the spectral gap can always be lower bounded by $\lambda \geq \cO(N^{-1} e^{-2\beta \overline{\epsilon}})$, where $N$ denotes the number of qubits in the stabilizer Hamiltonian and $\overline{\epsilon}$ is the generalized energy barrier that will be defined in eqn. (\ref{GenEng}). Furthermore, we show that in several cases the pre factor $N^{-1}$ can be removed, and we provide conditions addmitting an improvement of the lower bound to $\lambda \geq \cO( e^{-2 \beta \overline{\epsilon}} )$. We believe this to be the correct scaling of the spectral gap in the low - temperature limit, and are convinced that the prefactor $N^{-1}$, present in the general case, is an artifact of the method used to derive the generic bound. To illustrate the evaluation of this bound for a particular quantum memory we discuss the toric code, as well as one - dimensional models as a examples in section \ref{example}. We observe that the generalized energy barrier essentially corresponds to the largest energy barrier of the logical operators. The bound proves that although the existence of an energy barrier is not sufficient \cite{bravyi2013quantum,michnicki20143d}, it is certainly necessary.\\ 

The paper is organized as follows: First, in the remainder of this section we state the necessary background for Hamiltonians comprised of commuting Pauli operators and briefly introduce Davies generators for these Hamiltonians. Then in section \ref{sec:convergence} we provide an overview of the convergence analysis of Lindblad generators. The lower bound to the spectral gap for stabilizer Davies generators is derived in the section \ref{sec:lowerBnd}. This section contains the central technical contributions and states the lower bound in terms of a quantity which is very similar to the classical canonical paths bound derived by Jerrum and Sinclair. The final result which relates the spectral gap to the generalized energy barrier is presented in section \ref{sec:bound}. The reader only interested in the main result may skip to this section, where an intuitive description of the generalized energy barrier and examples are provided. 

\subsection{Preliminaries}
\label{sec:prelim}
Before we discuss the bounds on the equilibration times, we need to establish some background and notation. The Pauli group on the Hilbert space of $N$- qubits $\cH_N = \otimes_{i=1}^N \bC^{2}$ is defined as the group that results from the $N$-fold tensor product of the Pauli matrices $\1,\sigma^x,\sigma^y,\sigma^z$  so that  $\cP_{N} = \avr{i\1,X_1,Z_1,\ldots,X_N,Z_N}$. Here $X_i$ denotes the action of $\sigma^x$ on the $i$'th qubit and identity on the remaining $N-1$. Note, we will refer to the set of weight one Pauli operators as $W_1 = \{X_i,Y_i,Z_i\}_{i=1,\ldots,N}$. By weight one we refer to all Pauli operators that act non-trivially on only a single qubit. We consider Hamiltonians $H$ on $\cH_N$ that can be written as the sum of a set of Hermitian, commuting Pauli operators $\cG = \{ g_1,\ldots,g_M\} \subset \cP_N$, with $[g_i,g_j] = 0$ for all $i,j$. Together with the numbers $J_k \in \bR$ we can write the commuting Pauli Hamiltonian as
\be\label{CommPauli}
H   = -\sum_{k=1}^M J_k g_k.
\ee
The set $\cG$ is the generating set for the commuting subgroup $\cS = \avr{g_1,\ldots,g_M}$ of $\cP_N$.  This subgroup is referred to as the stabilizer group if it does not contain $-\1$. The stabilizer group encodes logical qubits in states that are stabilized, i.e. $s \ket{\psi} = \ket{\psi}$, by all $s \in \cS$, when $\cS$ is a strict subset of the centralizer ${\cal C}_{\cP_N}(\cS)$ of the stabilizer group. Pauli matrices that are contained in the set difference ${\cal C}_{\cP_N}(\cS) \backslash \cS$ are called logical operators and act non-trivially on the stabilized code space. The interested reader is referred to \cite{gottesman1997stabilizer} for a good introduction into stabilizer quantum codes. It is important to point out, that we do not assume that the set $\cG$ generates a stabilizer code in order to derive the thermalization time bound for $H$.  Although we will use the notation of stabilizer codes, the result holds for any commuting Pauli Hamiltonian. 

There is a natural way of identifying every element of the Pauli group $\cP_N$ with an element in $\bZ^2_2 \rtimes \bZ_2^{2N}$, where two bits in $\bZ^2_2$ are needed to encode the phase information \cite{gottesman1997stabilizer}. As we will be working with the Pauli algebra $\bC[\cP_N]$ we only associate vectors over the finite field $\bZ_2^{2N}$ with Pauli operators and drop the phase dependence from now on. This means, we consider for ${\alpha} \in \bZ_2^{2N}$, with ${\alpha} = ({\alpha}^x,{\alpha}^z)$ the projective representation $\sigma: \bZ_2^{2N} \raw \cP_N$, given by
\be
\sigma({\alpha}) = e^{i\frac{\pi}{2} \avr{{\alpha}^x,{\alpha}^z}_{\bZ}} \; {\bf X}^{{\alpha}^x} \; {\bf Z}^{{\alpha}^z},	
\ee
where we have defined the operators $ {\bf X}^{{\alpha}^x} = X_1^{\alpha_1^x} \otimes \ldots \otimes X_N^{\alpha_N^x} $ and for the Pauli $Z$ operators respectively $ {\bf Z}^{{\alpha}^z} = Z_1^{\alpha_1^z} \otimes \ldots \otimes Z_N^{\alpha_N^z} $. Note, that we will denote the addition modulo $2$ in $\bZ_2$ by $\oplus$ to distinguish it from the addition in $\bC$. We have that $X_i$ corresponds to a vector $(1,0)_i$, whereas $Y_i = (1,1)_i$ and $Z_i = (0,1)_i$ respectively. Furthermore, we have that for the product of $\sigma({\alpha}), \sigma({\beta})$ the following holds

\bq\label{PauliMult}
	\sigma({\alpha}) \sigma({\beta}) &=& e^{i\frac{\pi}{2} {Sp} ({\alpha},{\beta})_{\bZ}} \; \sigma({\alpha} \oplus {\beta}) \\
	\sigma({\alpha}) \sigma({\beta}) &=& \theta_{\alpha,\beta} \; \sigma({\beta}) \sigma({\alpha}) \Sp \mbox{with} \Sp \theta_{\alpha,\beta} = e^{i \pi {Sp}({\alpha},{\beta})_{\bZ}},
\eq 

where we define the symplectic product over $\bZ$ between ${\alpha}$ and ${\beta}$ as 

\bq
\mbox{Sp}({\alpha},{\beta})_{\bZ} = \left(\begin{array}{cc} {\alpha}^x & {\alpha}^z \end{array}\right)
 \left(\begin{array}{cc} 0 & -\1 \\ \1 & 0 \end{array}\right) \left(\begin{array}{c} {\beta}^x \\ {\beta}^z \end{array}\right).
\eq

The parameter $\theta_{\alpha,\beta} = \pm1$, depending on whether $\sigma(\alpha)$ and $\sigma(\beta)$ commute or anti-commute. The Pauli algebra $\bC[\cP_N]$ is defined via the span of the vectors $\bC[\cP_N] = \mbox{span}\{\oket{\gamma}\}_{\gamma \in \bZ_2^{2N}}$. For convenience we will use the representation $\oket{\gamma} \in \bC[\cP_N] \simeq \cH_N \otimes \cH_N$ given by

\bq\label{pauli_basis}
\oket{\gamma} = \sigma(\gamma) \otimes \1\ket{\Omega}, \Sp \mbox{with} \Sp \ket{\Omega} = \frac{1}{2^{N/2}} \sum_{k=0}^{2^N-1} \ket{k}\otimes\ket{k}.
\eq

Recall that the Pauli matrices form a complete orthonormal basis of the matrix algebra $\cM_{2^N}(\bC)$ with respect to the Hilbert-Schmidt scalar product. This implies immediately, that the vectors (\ref{pauli_basis}) form an orthonormal basis and the Pauli algebra is nothing but  $\cM_{2^N}(\bC)$.\\

It is convenient to introduce the binary matrix $G : \bZ_2^M \raw \bZ_2^{2N}$ to encode the generating set $\cG$. This matrix is of the form  

\bq 
G = \left(\begin{array}{c} G_X \no G_Z \end{array}\right), \Sp \mbox{where} \Sp  G_X , G_Z \in \cM_{N \times M}(\bZ_2). 
\eq

This binary matrix defines the elements of the generating set $\cG$ through it's column space. Since all elements in $\cG$ commute, we can identify every element $s \in \cS$ by an $M$ dimensional bit string $x \in \bZ_2^M$, through $s = \prod_{i=1}^M g_i^{x_i}$. This allows us to write any $s \in \cS$ in terms of the matrix $G$ by simply observing that 

\be
	s = \prod_{i=1}^M g_i^{x_i} = \sigma\left(\; G\; x \;\right).
\ee

When $\cG$ generates a stabilizer code, this matrix is often referred to as the code matrix. Quantum codes can for instance be obtained by choosing $G$ as the direct sum of two classical code matrices encoding the $X$ and $Z$ part independently \cite{Chuang}. We note that the generating set $\cG$ does not need to be independent, i.e. there may exist $x_1 \neq x_2 \in \bZ_2^M$ such that $Gx_1 = Gx_2$.  

This language allows for a very efficient representation of the spectrum of the Hamiltonian $H$. To every matrix $G$ we associate a matrix $E : \bZ_2^{2N} \raw \bZ_2^M$ referred to as parity check matrix  that can be obtained from $G$ through the identification $E = (G_Z^T , G_X^T)$. Since $G$ encodes a commuting set, we have that $EG = 0$. This matrix has the property that with the symplectic product as defined above we have for any $x \in \bZ_2^M$ and any $\alpha \in \bZ_2^{2N}$ that

\be\label{trick1}
	e^{i\pi Sp(\alpha,Gx)_\bZ} = e^{i\pi \avr{E \alpha,x}}.
\ee

Here, we denote by $\avr{a,b}$ the canonical inner product over $\bC^M$ and treat the vectors $E\alpha,x$ as belonging to this space. The parity check matrix $E$ plays an important role in coding theory, and allows for the detection of errors in a code. The image of $E$ will be referred to as the syndrome space and can be associated to the space of excitations of the Hamiltonian $H$. In essence given a Pauli matrix labeled by $\gamma$, the parity check matrix indicates the generators that anti-commute with this Pauli. These generators are then referred to as supporting an excitation. We will refer to the vector   

\be\label{linear:syndrome}
 e(\gamma) = E \gamma \in \bZ_2^M.  
\ee

as the syndrome of the Pauli $\gamma$. We denote by $e_k(\gamma) = [e(\gamma)]_k$ the $k$'th component of the syndrome vector.\\ 

The Pauli matrices $g_k$ have eigenvalues $\pm1$. The local projectors $\Pi_k(a_k) = 2^{-1}\left(\1 + e^{i\pi a_k}g_k\right)$, project onto the positive $a_k =0$ or negative $a_k = 1$ eigen space of the Pauli matrix $g_k$. Since all $g_k$ commute we can furthermore consider the product of all the local projectors $P(a) = \Pi_1(a_1) \ldots \Pi_M(a_M)$, for any $a \in \bZ_2^{M}$. Note that this projector can easily be expressed in terms of a $\bZ_2$ Fourier transform over the elements in $\cS$ through

\be\label{Fourier1}
	P(a) = \frac{1}{2^M} \sum_{x \in \bZ_2^{M}} e^{i\pi\avr{a,x}}\sigma(G x).
\ee

The inverse is naturally given by $\sigma(G x) = \sum_{a} e^{i\pi\avr{a,x}} P(a)$ and one can immediately verify that $\sum_a P(a) = \1$.  Since we have already stated that the set $\cG$ is not necessarily independent, we also observe that there may be an $a \in \bZ_2^M$ for which $P(a) = 0$.  The $a$ for which $P(a)$ does not vanish coincides with the image of the parity check matrix $E$ and will be referred to as being in the syndrome space of $\cG$. The projectors $P(a)$ satisfy an important identity when conjugated by Pauli operators. It can be verified by making use of the identity in eqn. (\ref{trick1}) and the Fourier expansion eqn. (\ref{Fourier1}), that the projectors satisfy 

\bq \label{Pauli-com}
	\sigma(\alpha) P(a) \sigma(\alpha) &=& P(a \oplus E \alpha).
\eq

Since this addition $a \oplus E \alpha$ of syndromes will appear frequently we will write as shorthand notation  

\be \label{notation:syn}
a^\alpha = a \oplus e(\alpha).
\ee  

The projectors $P(a)$ can now be used to diagonalize the Hamiltonian so that we can write 
\bq
	H = \sum_{a \in \bZ_2^M} \epsilon_a P(a)\Sp \mbox{with eigenvalues} \Sp \epsilon_a = -\sum_k J_k (-1)^{a_k}.
\eq

From this particular form, it is straight forward to compute the Gibbs distribution $\rho = Z^{-1} \exp(-\beta H)$ and we obtain that 

\be\label{Gibbs-rep}
\rho =  \frac{1}{Z} \sum_{a} e^{-\beta \epsilon_a} P(a) = \sum_{a} \rho_a P(a).
\ee 

\subsection{Davies generator}

We will describe the thermalization of the system in terms of a Davies generator. This generator has assumed the role of a bona fide standard model in the description of thermalization in quantum memories. The Lindblad master equation arrises from the weak coupling limit of the system to a thermal heat bath. For a microscopic derivation, the reader is referred to \cite{Davis,Davis2,alicki2007quantum,Breuer}. We will consider the generator as given and will not focus on its derivation. The physical picture is the following: We assume that the system and bath evolve together under the Hamiltonian $H_{tot} = H + H_B + H_I$, where $H_B$ denotes the bath Hamiltonian, which we will not specify here. The Bath is in a Gibbs state with respect to $H_B$ at some fixed temperature $\beta$. We assume a weak interaction between system and bath given by $ H_I = \sum_{\alpha} S^\alpha \otimes B^\alpha.$ Here $S^\alpha$ is a Hermitian operator that acts only on $\cH_N$, whereas $B^\alpha$ is some Hermitian bath operator. After tracing out the bath degrees of freedom and a complex sequence of approximations one is left with a Lindblad master equation of the form $\partial_t \sigma_t = -i[H_{eff}, \sigma_t] + \cL_\beta^*(\sigma_t)$. It can be shown, that the effective Hamiltonian term $[H_{eff},\sigma_t]$ does not contribute to the spectral gap \cite{temme2013lower,alicki2009thermalization} and we therefore neglect this term here. We will therefore only refer to the term $\cL_\beta$ as Davies generator for convenience. The generator is given by

\be\label{Davies}
	\cL_\beta(f) = \sum_{\alpha \in W_1} \sum_{\omega} h^\alpha(\omega)\left({S_\omega^\alpha}^\dag f {S_\omega^\alpha}  - \half\left\{{S_\omega^\alpha}^\dag {S_\omega^\alpha},f\right\}\right).
\ee

For our model, we make the assumption that the system couples to the bath via single qubit Pauli operators, $  S^\alpha = \sigma(\alpha) \in  W_1 = \{X_i,Y_i,Z_i\}_{i=1,\ldots,N} $. The second sum over $\omega$ is a sum over all Bohr frequencies of the commuting Pauli Hamiltonian $H$. A Bohr frequency $\omega = \epsilon_a - \epsilon_b$ is an eigenvalue difference of the the Hamiltonian. The operators $S_\omega^\alpha$ are obtained from the coupling operators through the Fourier expansion of $\exp(i H t) S^\alpha \exp(-i Ht) = \sum_{\omega} S_\omega^\alpha e^{i \omega t}$.  Since we can diagonalize the Hamiltonian $H$ and the individual summands commute we can compute the time evolution of $S^\alpha$ and read off the components 
\bq\label{LBop}
	{S_\omega^\alpha} = \sum_{a} \delta[\omega^\alpha(a) - \omega]  \sigma(\alpha) P(a), \Sp \mbox{with} \Sp \delta[x] = \left \{ \begin{array}{c} \;\; 1 \Sp : \Sp x = 0 \\ 0 \Sp : \Sp \mbox{else}. \end{array} \right.
\eq
We have defined $\omega^\alpha(a) = \epsilon_a -\epsilon_{a^\alpha}$. Due to the particularly simple form of the eigenvalues, the Bohr frequency can be evaluated as  

\be 
\omega^\alpha(a) = -2\sum_{k=1}^M J_k e_k(\alpha) (-1)^{a_k}.
\ee

Note that we consider the binary variables $e_k(\alpha)$ as $0,1$ valued integers and use the natural addition. The bath temperature is encoded in the {\it transition rates} $h^\alpha(\omega)$. This function is obtained from the Fourier transform of the autocorrelation of the bath operator $B^\alpha(t) = \exp(i H_B t) B^\alpha \exp(-i H_B t)$ with respect to the bath's Gibbs state at inverse temperature $\beta$. The specific form of the transition rates depends of course on the particular choice of bath model \cite{weiss1999quantum}. However, the only property which is relevant for our derivation is that the transition rates satisfy the KMS condition \cite{kossakowski1977quantum}

\be \label{KMS}
 h^\alpha(-\omega) = h^\alpha(\omega)e^{-\beta \omega},
\ee

to ensure detailed balance, c.f. definition \ref{DetailedBalance}. Moreover, we assume that the functions are positive and bounded by $c \leq h^\alpha(\omega) \leq C$, where $c,C>0$ are constants independent of $N$. In particular we will assume that the lower bound behaves as $c \sim e^{-\beta \Delta}$, where $\Delta$ is the gap of the Hamiltonian (\ref{CommPauli}). The coupling operators  $S^\alpha \in W_1$ ensure that $\cL_\beta$ has a unique full rank stationary state $\rho > 0$ for which $\cL^*_\beta(\rho) = 0$, since $W_1$ generates the full algebra \cite{Spohn,Wielandt}. Furthermore, the detailed balance of $\cL_\beta$ with respect to the Gibbs state of $H$ implies that the unique fixed point of this map is given by $\rho = Z^{-1}\exp(-\beta H)$. We therefore have that $\cL_\beta$ is a map that converges to the thermal state of the Hamiltonian $H$. The Davies generator can therefore be seen as a physically motivated generalization of Glauber dynamics to quantum systems \cite{Martinelli,Martinelli2}.  

\section{The Poincare inequality and Convergence bounds}
\label{sec:convergence}

We are interested in the derivation of convergence time bounds for the Davies generator (\ref{Davies}) defined in the previous section. In order to analyze the convergence of density matrices we will work with the trace norm $\| A \|_{tr} = \tr{\sqrt{A^\dag A}}$ to determine the distance from the steady state. This norm is the natural non-commutative generalization of the total variation distance \cite{Fuchs}. Let us denote the steady state of the Davies generator by $\rho$.  We will define the {\it convergence time}, or so-called {\it mixing time}, $t_{mix}(\epsilon)$ as the time the semi-group $\cL_\beta$ needs to be $\epsilon$-close to its stationary distribution for all initial states $\sigma_0$. 

\be
	t_{mix}(\epsilon) = \min \left\{ t \;\;\big {|}\;\;  t' > t  \Sp \mbox{we have} \Sp \|e^{\cL^*_\beta t'}(\sigma_0) - \rho\|_{tr} \leq \epsilon \;\;\forall \;\sigma_0 \right\}.
\ee

The mixing time gives a valid estimate for the thermalization time of the quantum system. Moreover, this time also provides an upper bound to the time information can be encoded in the system. Once the system has become thermal it has lost all information of its initial configuration. Note that the system's ability to store quantum information may be lost before the Hamiltonian starts to thermalize. Hence, the mixing time bounds only what is referred to as the classical memory time. This time is of course a natural upper bound to the life time of a quantum memory. To find appropriate upper bounds to the mixing time, we take an approach that was developed in \cite{sinclair1989approximate,diaconis1991geometric,fill1991eigenvalue} and generalized to quantum mechanical semi-groups in \cite{chi2}. We need to have access to the spectral gap $\lambda$ of the generator $\cL$. Here, the spectrum of the map $\cL$ is understood in terms of the matrix representation of $\cL$ on the vector space $\cM_{2^N}(\bC) \simeq \bC^{2^N \times 2^N}$. The spectral gap $\lambda$  of $\cL$ will be introduced properly in lemma \ref{spectGapVar-lem}. In \cite{chi2} the following exponentially decaying bound, which holds for any $\cL$ was proven.

\begin{theorem}\label{Mixingtimebound}
Let $\cL:\cM_d\rightarrow\cM_d$ be a Liouvillian with stationary state $\rho$ and spectral gap $\lambda$ Then the following trace norm convergence bound holds: 
\be
	\left\| \sigma_t - \rho \right \|_{tr} \leq \sqrt{\|{\rho}^{-1}\|}e^{-\lambda t}.
\ee  
Here $\|{\rho}^{-1}\|$ denotes the inverse of the smallest eigenvalue of the stationary state, and $\sigma_t = e^{t \cL^*}( \sigma_0)$ for initial state $\sigma_0$.
\end{theorem}

The convergence result of theorem \ref{Mixingtimebound} provides a simple upper bound on the mixing time.  Recall that we consider thermalizing semi-groups, for which the fixed point is always given by the Gibbs distribution for which $\|\rho^{-1}\| \leq \exp(\mbox{const} \beta N)$. Let us now choose a fixed $\epsilon = e^{-1/2}$ for which $t_{mix} \equiv t_{mix}(e^{-1/2})$. One can easily rearrange the upper bound to find that we can choose 

\be\label{bound-on-t}
	t_{mix} \leq \cO(\beta N \lambda^{-1}).
\ee

The bound on the mixing time derived from the spectral gap scales at least linearly in the the system size $N$, even when the spectral gap is a constant independent of $N$. Other approaches to bounding the mixing time exist, which can yield bounds that can scale as $\cO(\log(N))$. These bounds are based on logarithmic Sobolev inequalites \cite{gross1975logarithmic,diaconis1996logarithmic,kastoryano2013quantum} which are more challenging to prove in general \cite{olkiewicz1999hypercontractivity,temme2014hypercontractivity}. 

The spectral properties of the generator (\ref{Davies}) can best be understood when working with an inner product that is weighted with respect to some full rank reference state $\rho > 0$.  This reference state is typically chosen as the fixed point of the Liouvillian, i.e. the Gibbs state. We furthermore introduce the variance and the Dirichlet form, which will play an important role in the spectral analysis of the semi-group.  

\begin{definition}
Given a full rank state $\rho$ and a Liovillian $\cL$,  we define the following quadratic forms on $\cM_{2^N}(\bC)$:

\begin{enumerate}
\item  The $\rho$-weighted non-commutative \textbf{inner product}  for all $f,g \in \cM_{2^N}(\bC)$: 
\be\label{inner-prod}
\avr{f,g}_\rho = \tr{\rho f^\dag g}.
\ee

\item The \textbf{variance} of $f \in \cM_{2^N}(\bC)$ with respect to $\rho$:
\be\label{Variance}
	\Var_\rho(f,f) = \tr{\rho f^\dag f} - |\tr{\rho f}|^2.
\ee

\item The \textbf{Dirichlet form} of $\cL$ with respect to $\rho$:
\be\label{Dirichlet}
	\cE(f,f) =  - \avr{f,\cL(f)}_\rho = - \tr{\rho f^\dag \cL(f)}.
\ee
\end{enumerate}
\end{definition} 

These quantities give convenient access to the spectral properties of the Davies generator. Lindblad generators in general may have a complex spectrum, which makes it necessary to be more careful in the definition of the spectral gap \cite{chi2}. For Davies generators, however, this is not the case since this map becomes Hermitian with respect to the previously defined $\rho$ weighted inner product. We will refer to this property as detailed balance and give its formal definition below. 

\begin{definition}\label{DetailedBalance} We say a Liouvillian $\cL$ satisfies \textbf{detailed balance} (or is \textbf{reversible}) with respect to the state $\rho > 0$, if   
\be
\avr{f,\cL(g)}_\rho = \avr{\cL(f),g}_\rho
\ee
for all $f,g \in \cM_{2^N}(\bC)$.
\end{definition}

It follows from the KMS condition discussed previously in eqn. (\ref{KMS})  that the Davies generator is reversible \cite{db1} with respect to the Gibbs distribution. This was already shown in the seminal work by Davies \cite{Davis,Davis2}. Detailed balance immediately implies two things: First, that the spectrum of $\cL_\beta$ is real. Second, as can be verified easily, reversibility ensures that the state $\rho = Z^{-1}\exp(-\beta H)$ is a fixed point of the Liouvillian \cite{chi2}. Moreover, since we consider the case where the system is coupled via all single qubits Pauli $W_1 = \{X_i,Y_i,Z_i\}_{i=1,\ldots,N}$ operators to the bath, we automatically have that the Gibbs state is the unique fixed point \cite{Spohn}. We are now ready to find a convenient variational expression for the spectral gap of the Davies generator. The following lemma was proved in \cite{chi2}.

\begin{lemma}\label{spectGapVar-lem} The \textbf{spectral gap} of a primitive Liouvillian $\cL: \cM_{2^N}(\bC) \rightarrow \cM_{2^N}(\bC)$ with stationary state $\rho$ is given by the variational expression
\bq \label{spectGapVar}
	\lambda  =   \min_{f \in \cM_{2^N}}  \frac{\cE(f,f)}{\Var_{\rho}(f,f)}.
\eq  
Note that $f \in \cM_{2^N}(\bC)$ in the optimization can be chosen as a Hermitian matrix.
\end{lemma}

This  lemma  leads to a very useful inequality referred to as the {\it Poincare inequality}. It is clear that  the problem of finding good lower bounds to the spectral gap can be rephrased as the problem of finding a constant $\lambda$ so that the inequality 

\be \label{Poincare}
\lambda \Var_{\rho}(f,f) \leq \cE(f,f)
\ee 

is satisfied for all Hermitian $f$. This inequality will be the starting point to prove spectral gap lower bounds for the Davies generator. \\

Lower bounds to $\lambda$ in the Poincare inequality can be found for instance by expressing the inequality for the two quadratic forms in terms of a matrix inequality. We make use of the vectorization of $f$ through $\oket{f} = f \otimes \1 \ket{\Omega}$ as discussed in the previous section \ref{sec:prelim}. Both the quadratic forms can be written as 
\be
\Var_{\rho}(f,f) = \obra{f} \hat{\cV} \oket{f} \Sp \mbox{and} \Sp \cE(f,f) = \obra{f} \hat{\cE} \oket{f}.
\ee
The matrices  $\hat{\cV}$ and $\hat{\cE}$ will be explicitly given in section \ref{sec:lowerBnd}. The Poincare inequality (\ref{Poincare}) is then trivially equivalent to a positive semi-definite matrix inequality, where we now want to find the smallest $\tau \in \bR$ such that the following holds, 
\be
	\tau \hat{\cE} - \hat{\cV} \geq 0.
\ee
It is clear that this optimal $\tau$, which is often also referred to as support number,  is related to the spectral gap via $\tau = \lambda^{-1}$. Any upper bound on $\tau$ will immediately constitute a lower bound on the spectral gap $\lambda$. Note, that $\tau$  is well defined even for singular matrices, as long as $\ker(\hat{\cE}) \subset \ker(\hat{\cV})$. This will be the case here, since $\cL$ is ergodic so that both maps have the same kernel given by the identity. 

A very useful lemma to finding bounds on $\tau$ was developed in \cite{boman2003support,chen2005obtaining}. It is possible to express $\tau$ as the constrained minimization over a certain matrix factorization. We therefore have that any factorization that satisfies the constraints gives rise to a valid upper bound on the support number. This is expressed in a lemma first proved in \cite{boman2003support}.

\begin{lemma}\label{lem:W-bound}
Let $\hat{\cE},\hat{\cV}$ be positive semi-definite with a decomposition $\hat{\cE} = A A^\dag$  and $\hat{\cV} = B B^\dag$. then the minimal $\tau$ for which the matrix $\tau \hat{\cE} - \hat{\cV}$ is positive semi-definite is given by
\be
	\tau = \min_{W} \| W \|^2 \Sp \mbox{subject to} \;\;  A W =B.
\ee
Here, $\|W\|$ denotes the operator norm, i.e largest singular values, of $W$.
\end{lemma}

The direct evaluation of the  operator $\| \cdot\|$ norm does at first appear to be just as challenging as the original problem. However, since we are only trying to find upper bounds on $\tau$ suitable norm inequalities will suffice.  Once such a factorization is found, several different norm bounds can be used to yield different lower bounds to the spectral gap. One common choice is for instance given by Schur's bound \cite{Bhatia} on the operator norm  $\|W\|^2 \leq \|W\|_{\infty}\|W\|_1$, where $\|W\|_1$ and $\|W\|_{\infty}$ denote the maximal row and column sum of $W$ respectively. The bound on the operator norm which will be most relevant to us has been introduced in \cite{chen2005obtaining}, since it does yield a lower bound to the spectral gap which is very similar to the canonical paths bound for classical Markov chains given in \cite{sinclair1992improved,sinclair1989approximate,diaconis1991geometric,fill1991eigenvalue}. 

\begin{lemma}\label{normBND}
Let $W \in \cM_{K,M}(\bC)$ denote a complex rectangular matrix $W = \sum_{k=1}^K\sum_{m} W_{k,m} \ket{k}\bra{m}$ with row vectors $\ket{w_k} = \sum_{m=1}^M W_{k,m} \ket{m}$, then the operator norm 
of $W$ is bounded by
\be
\| W \|^2 \leq \max_{m} \sum_{k \; : \; W_{km} \neq 0} \|\ket{w_k}\|_2^2.	
\ee
\end{lemma}

\proof{ We follow the proof in \cite{chen2005obtaining}. Given the matrix $W$, suppose we could find an isometry $S$ with $SS^\dag = \1$ and a matrix $\tilde{W}$ such that $W = S \tilde{W}$, then we can bound $\| W \|^2  \leq \| S \|^2 \| \tilde{W} \|^2  = \| \tilde{W} \|^2$, since the operator norm of $S$ is bounded by unity. Moreover, if we can find a $\tilde{W} = \sum_{k'=1}^{K'} \sum_{m=1}^M \tilde{W}_{k' m} \ket{k'} \bra{m}$ such that it's columns $\ket{\tilde{w}_m} = \sum_{k'}  \tilde{W}_{k'm} \ket{k'}$ are orthogonal, we have that 
\be\label{sd9fa0}
	\| W \|^2 \leq \max_{m}  \| \ket{\tilde{w}_m} \|^2_2.
\ee
Now, consider the matrix pair
\bq
S &=& \sum_{k=1}^K \sum_{m=1}^M \frac{W_{k,m}}{\|\ket{w_k}\|_2} \ket{k}\bra{k}\otimes \bra{m} \Sp \mbox{and} \no
 \tilde{W} &=& \sum_{k=1}^{K} \sum_{m=1}^M \|\ket{w_k}\|_2(1 - \delta[W_{k,m}]) \ket{k}\otimes \ket{m} \bra{m}.
\eq
One can easily see that the constraints on $S$ and $\tilde{W}$ are met so that $\| \ket{\tilde{w}_m} \|^2_2 = \sum_{k \; : \; W_{km} \neq 0} \|\ket{w_k}\|_2^2$ and by (\ref{sd9fa0}) the bound as stated in the lemma holds.\\ \qed }

In order to derive the spectral gap bound, we now proceed as follows: First we find suitable matrix representations for $\hat{\cE}$ and $\hat{\cV}$, then we try to find a factorization in terms of a matrix triple $[A,B,W]$ as given in lemma \ref{lem:W-bound}. An upper bound on the constant $\tau$ is then obtained by applying the norm bound from lemma \ref{normBND}.

\section{Lower bound to the spectral gap}
\label{sec:lowerBnd}

The central task is now to find a suitable upper bound on the support number of the matrix pair that stems from the Dirichlet form and the variance. We do so by first finding the matrices that constitute the quadratic forms and then by expressing this matrix in a suitable basis. It turns out, that the most natural operator basis to work with is given by the Pauli matrices considered earlier. Since the stabilizer group acts as a sub group in this algebra, we will find that both the variance, as well as the Dirichlet form can be expressed efficiently. 

\subsection{Matrix Representations of the quadratic forms}

As discussed in the previous section, we now proceed to derive the matrices that give rise to the quadratic forms $\cE(f,f) = \obra{f}{\hat{\cE}}\oket{f}$ and $\Var_{\rho}(f,f) = \obra{f} {\hat{\cV}} \oket{f}$. We choose the Pauli matrices as a basis of $\cM_{2^N}(\bC)$. 

Recall that $\cS$ is a subgroup of the full Pauli group $\cP_N$, we can therefore consider the right cosets of $\cS$ in $\cP_N$. For each coset we can define a suitable coset algebra, which is naturally a subspace of $\bC[\cP_N] \simeq \cM_{2^N}(\bC)$. The full algebra can then be decomposed in terms of its cosets. This is a decomposition which will turn out to be useful in the following. Assume we are given some representative $\sigma(\gamma_0) \in \cP_N$, then the right coset $\cS\sigma(\gamma_0)$, for which we will write $[\gamma_0]$ is spanned by the Pauli matrices $\sigma(G x)\sigma(\gamma_0)$ for $x \in \bZ_2^M$. So that the coset algebra is spanned by the vectors
\be 
	\bC[\gamma_0] = \mbox{span}\left\{\oket{G x \oplus \gamma_0}\right\}_{x \in \bZ^{M}_2}.
\ee

Moreover, it will become important later to also consider the dual algebra of the coset which is obtained by a $\bZ_2^M$ Fourier transform. The dual algebra of each coset $[\gamma_0]$ given by

\be
\bC{[\gamma_0]}^* =  \mbox{span}\Big{\{} \oket{a}_{\gamma_0} \Big{\}}_{a \in \bZ_2^{M}}
\ee

is spanned by the vectors

\be\label{dual}
	\oket{a}_{\gamma_0} = \frac{1}{2^{M/2}} \sum_x e^{i\pi \avr{x,a}} e^{i \frac{\pi}{2}Sp(Gx,\gamma_0)} \oket{Gx \oplus \gamma_0}. 
\ee

These vectors form an orthonormal basis. Recall that, depending on the generating set $\cG$, for some $a$ the projection operators $P(a)$ can  vanish. This pathology carries over to the vectors $\oket{a}_{\gamma_0}$. This however, is not relevant for our analysis here, since we can always interpret these $a$ values as being omitted in the sum so that we sum only over legitimate syndromes of $\cG$.  We  now consider the decomposition of $\hat{\cE}$ and $\hat{\cV}$ in terms of this basis. 

\begin{lemma}
The matrix $\hat{\cE}$ is block diagonal over the right cosets $\bC[\gamma_0]$ of the subgroup $\cS$ with representatives $\gamma_0 \in \bZ_2^{2N}$  in the full Pauli group $\cP_N$, 
\bq
\hat{\cE}  = \bigoplus_{[\gamma_0]} \hat{\cE}_{\gamma_0},
\eq
where every $\hat{\cE}_{\gamma_0}$ is only supported on  $\bC[\gamma_0]$. Moreover, we can write each block as \be \hat{\cE}_{\gamma_0} = \sum_{\alpha \in W_1} \sum_{a} \hat{\cE}_{\gamma_0}^\alpha(a),\ee with
\bq \label{Eq:DirichletSummand}
\hat{\cE}_{\gamma_0}^\alpha(a) =\half\left(h_{aa}^\alpha + h_{a^{\gamma_0} a^{\gamma_0}}^\alpha  \right) \rho_a  \oproj{a}_{\gamma_0} - h_{a  a^{\gamma_0}}^\alpha \rho_a  \theta_{\alpha,\gamma_0} \oket{a}\obra{a^\alpha}_{\gamma_0}.
\eq
Where we have that $h^\alpha_{a,b} = h^\alpha(\omega^\alpha(a))\delta[\omega^\alpha(a)  - \omega^\alpha(b)]$.
\end{lemma}

\proof{ The Davies generator can be split into a sum over the individual coupling operators as $\cL_\beta(f) = \sum_{\alpha \in W_1} \cL_\beta^\alpha(f)$. The individual $\cL_\beta^\alpha(f)$ are obtained from Eqn. (\ref{Davies}) by substitution of $S^\alpha_\omega$  as in given in  Eqn. (\ref{LBop}). A summation over all values of $\omega$ then yields 

\bq
 \cL^\alpha_\beta(f) &=& \sum_{\omega} h^\alpha(\omega)\left({S_\omega^\alpha}^\dag f {S_\omega^\alpha}  - \half\left\{{S_\omega^\alpha}^\dag {S_\omega^\alpha},f\right\}\right) \no
 		               &=& \sum_{ab} h^\alpha_{ab} \left( P(a) \sigma(\alpha) f \sigma(\alpha) P(b) - \frac{\delta_{a,b}}{2} \{P(a),f\}_+\right).
\eq

We want to find a matrix that represents the Dirichlet form $\cE(f,f)$. This means that we need to find a matrix $\hat{\cE}^\alpha$  for every summand $\alpha \in W_1$ so that $-\tr{ \rho \cL^\alpha_\beta(f) f} = \obra{f}\hat{\cE}^\alpha\oket{f}$ for any $\oket{f} \in \bC[\cP_N]$. Note that we have made use of detailed balance here. Let us therefore consider the action of this map on some matrix $f \in \cM_{2^N}(\bC)$ for which we can then write $\cE^\alpha(f) = - \rho\cL^\alpha_\beta(f)$. The Gibbs state can be written as $\rho = \sum_a \rho_a P(a)$, so a direct substitution yields the result

\be
\cE^\alpha(f) =   \sum_{a,b} \half\left(h^\alpha_{aa} + h^\alpha_{bb}\right)\rho_aP(a) f P(b)  - h^\alpha_{a,b}\rho_a P(a)\sigma(\alpha) f \sigma(\alpha) P(b).
\ee

We will work in the Pauli basis, so that we need to understand the action of $\cE_\alpha$, on any $2^{-N/2}\sigma(\gamma)$. With the commutation relation (\ref{Pauli-com}) between the projectors $P(a)$ and any Pauli we have that $P(a)\sigma(\gamma)P(b) = P(a) \delta_{a,b \oplus e(\gamma)} \sigma(\gamma)$. Furthermore we can write for any $\sigma(\alpha) \sigma(\gamma) \sigma(\alpha)= \theta_{\alpha,\gamma} \sigma(\gamma)$, where $\theta_{\alpha,\gamma} = \pm 1$ was introduced in eqn. (\ref{PauliMult}). We obtain 

\bq
\cE^\alpha(\sigma(\gamma)) &=& \sum_{a} \left[ \half\left(h_{aa}^\alpha + h_{a^\gamma a^\gamma}^\alpha  \right)\rho_a 
					           -h^\alpha_{a,a^\gamma}\rho_a \; \theta_{\alpha,\gamma} \right] \;P(a) \sigma(\gamma).	 \nonumber
\eq

Recall, that $P(a) = \sum_x 2^{-M} e^{i\pi \avr{a,x}} \sigma(G x)$, so that this substitution yields the double sum

\bq
\cE^\alpha(\sigma(\gamma)) &=& \frac{1}{2^{M}} \sum_{a,x} \left[\half\left(h_{aa}^\alpha + h_{a^\gamma a^\gamma}^\alpha  \right)  - \theta_{\alpha,\gamma} h_{aa^\gamma}^\alpha\right] \rho_a e^{i\pi \avr{a,x}} \sigma(G x) \sigma(\gamma). \nonumber
\eq

Since we now understand the action of $\cE^\alpha$ on the Pauli matrices $2^{-N/2} \sigma(\gamma)$, we can express the matrix now in terms of the operator basis elements $\oket{\gamma}$.  The multiplication rule for the Pauli matrices was given in (\ref{PauliMult}). Since all Paulis are orthogonal, we can write

\bq\label{dude}
\hat{\cE}^\alpha = \frac{1}{2^M} \sum_\gamma \sum_{a,x} \left[\half\left(h_{aa}^\alpha + h_{a^\gamma a^\gamma}^\alpha  \right)  - \theta_{\alpha,\gamma} h_{aa^\gamma}^\alpha\right] \; \rho_a e^{i\pi \avr{a,x}} e^{i\frac{\pi}{2} {Sp}(Gx,\gamma)} \oket{ G x \oplus \gamma}\obra{\gamma}.
\eq

To simplify the notation in the following we write

\bq\label{Dirichlet_functions} 
E^1_{\alpha,\gamma}(a) = \half\left(h_{aa}^\alpha + h_{a^\gamma a^\gamma}^\alpha  \right)\rho_a \Sp \mbox{and} \Sp
E^2_{\alpha,\gamma}(a) =  h_{aa^\gamma}\rho_a.
\eq

We observe that since both $\rho_a$ and $h(\omega^\alpha(a))$ only depend on elements of the syndrome space, we have that both $E^1_{\alpha,\gamma}(a)$ and $E^2_{\alpha,\gamma}(a)$ only depend on the syndromes $e(\alpha)$ and $e(\gamma)$ and not on the specific Pauli's $\alpha,\gamma$ themselves. Since we have that $EG = 0$ it can be inferred that the syndromes of two Pauli operators agree $e(\gamma_1) = e(\gamma_2)$, if the operators are related by an element in $\cS$. Hence the functions $E^{1/2}_{\alpha,\gamma}(a)$ are in fact constant in $\gamma$ over the cosets. Moreover, we can decompose the full Pauli group $\cP_N$ in terms of its right cosets $\cP_N = \cup_{i} [\gamma_i]$. Hence we can choose some representative $\gamma_0 \in \bZ_2^{2N} / \bZ_2^{M} $ and $y \in \bZ_2^{M}$ so that any Pauli can be written as $\gamma = G y \oplus \gamma_0$. We can therefore write
 
\bq
\hat{\cE}^\alpha &=& \frac{1}{2^M} \sum_{[\gamma_0]} \sum_a \sum_{x,y} \left[ E^1_{\alpha,\gamma_0}(a) -  E^2_{\alpha,\gamma_0}(a)\theta_{\alpha,Gy + \gamma_0} \right] \no 
&& \times e^{i\pi \avr{a,x}} e^{i\frac{\pi}{2}  Sp(Gx,Gy + \gamma_0)} \oket{G(x\oplus y) \oplus \gamma_0}\obra{G y \oplus \gamma_0}.
\eq

We define the matrix the diagonal matrix  

\be
\hat{\Theta}_{\alpha,\gamma_0} = \sum_x  \theta_{\alpha,Gx + \gamma_0} \oket{Gx \oplus \gamma_0}\obra{Gx \oplus \gamma_0},
\ee 

Furthermore, we define two bit strings  $x_1 = x \oplus y$ and $x_2 = y$, for which then $ \exp{(i\pi \avr{x,a})} = \exp \left(i\pi\avr{x_1,a} -i\pi\avr{x_2,a}\right)$.  Moreover, since $Sp(Gx, Gy) = 0$, which from the fact that all elements in $\cS$ commute, we have that $Sp(Gx,Gy + \gamma_0) = Sp(Gx_1, \gamma_0) - Sp(Gx_2, \gamma_0)$. We can write with with the dual basis $\oket{a}_{\gamma_0}$ as defined in (\ref{dual})

\bq
 \hat{\cE}^\alpha &=&  \sum_{[\gamma_0]} \sum_a E^1_{\alpha,\gamma_0}(a) \oproj{a}_{\gamma_0}  -  E^2_{\alpha,\gamma_0}(a) \oproj{a}_{\gamma_0} \hat{\Theta}_\alpha.
\eq

Note that, $\oproj{a}_{\gamma_0}$ is only supported on $\bC[\gamma_0]$. Hence, we have that for every $[\gamma_0]$ the matrix can be decomposed into disjoined blocks and we can write for
$\hat{\cE}^\alpha = \oplus_{[\gamma_0]} \hat{\cE}^\alpha_{\gamma_0}$, where the blocks are given by

\be\label{someMatrix123}
\hat{\cE}^\alpha_{\gamma_0} =  \sum_a  \oproj{a}_{\gamma_0}\left(E^1_{\alpha,\gamma_0}(a) -  E^2_{\alpha,\gamma_0}(a)\hat{\Theta}_\alpha\right).
\ee

Let us now look at $\hat{\Theta}_{\alpha,\gamma_0}$, this map was originally diagonal in the Pauli basis. However in the dual basis $\oket{a}_{\gamma_0}$ we have that, due to the identity (\ref{trick1}) and an application of the $\bZ_2^M$ Fourier transform, the matrix can be written as 

\be
\hat{\Theta}_\alpha = \theta_{\alpha,\gamma_0} \sum_a \oket{a}\obra{a^\alpha}_{\gamma_0}.
\ee  

Applying this matrix to $\oproj{a}_{\gamma_0}$ in (\ref{someMatrix123}) and taking the sum over  $\alpha \in W_1$, yields the decomposition of the matrix $\hat{\cE}$ as stated in the lemma.  \qed}

\vspace{0.5cm}
{\bf Remark:} In the derivation of the matrix $\hat{\cE}^\alpha_{\gamma_0}$ we have made the choice of a particular representative $\gamma_0$ for the coset. Here we will see, that the matrices are in fact independent of the representative. Any other $\gamma_1$ in the same coset is related to $\gamma_0$ by   $\gamma_1 = G x^* \oplus \gamma_0$. If we consider the dual vectors $\oket{a}_{\gamma_1}$, we can see that these are related to the ones defined by $\gamma_0$ by  $\oket{a}_{\gamma_1} = \exp(i\pi \avr{a, x^*}) \oket{a}_{\gamma_0}$. This follows from expanding $\oket{a}_{\gamma_1}$ in the basis $\{\oket{Gx + \gamma_1}\}$ and using the identity (\ref{trick1}). Since the vector only changes by a phase, the projectors $\oproj{a}_{\gamma_1} = \oproj{a}_{\gamma_0}$ are in fact identical. However, the matrix unit of the new representative changes according to $\oket{a}\obra{a^\alpha}_{\gamma_1} = \exp(i\pi\avr{x^* e(\alpha)})\oket{a}\obra{a^\alpha}_{\gamma_0} $. This is nevertheless consistent with the phase $\theta_{\alpha,\gamma_1}$ in the equation. Since $\exp(i \pi Sp(\alpha,Gx^* + \gamma_0)) = \exp(i \pi \avr{e(\alpha) , x^*} + Sp(\alpha,\gamma_0))$ we have that $\theta_{\alpha,\gamma_1} = \exp(i\pi \avr{x^*,e(\alpha)}) \theta_{\alpha,\gamma_0}$ canceling the phase from the matrix unit.  This  leads to the observation that if $\gamma_1$ and $\gamma_0$ are related as stated above, i.e. they belong to the same coset, we have that $\hat{\cE}^\alpha_{\gamma_1} = \hat{\cE}^\alpha_{\gamma_0}$. \\

Furthermore it is easy to see that the matrix $\hat{\cE}^\alpha_\gamma$ is Hermitian, which is a direct consequence of the KMS condition $h(-\omega^\alpha(a)) = \exp(-\beta \omega^\alpha(a))h(\omega^\alpha(a))$. This condition ensures that $E^2_{\alpha,\gamma}(a) = E^2_{\alpha,\gamma}(a^\alpha)$. One can therefore verify easily by simple Hermitian conjugation and a substitution of the labels according to $a \raw a^\alpha$ that $\hat{\cE}^\alpha_{\gamma_0} = {{\cE}^\alpha}_{\gamma_0}^\dag$.\\
 
We now need to see whether it is in fact possible to find a decomposition of $\hat{\cV}$ that is similar to the one of $\hat{\cE}$. If the two matrices are not too different form each other, we stand a good chance to factor them according to lemma \ref{lem:W-bound} and bound the spectral gap this way. Indeed, it turns out that the matrix $\hat{\cV}$ obeys the same block diagonal structure and is in many ways rather similar to $\hat{\cE}$. 

\begin{lemma}
The matrix $\hat{\cV}$ is block diagonal over the left cosets  $[\gamma_0]$ of the stabilizer group $S$ in the Pauli - group $\cP_N$. This matrix can be written as
\bq
\hat{\cV}  = \bigoplus_{[\gamma_0]} \hat{\cV}_{\gamma_0}.
\eq
Here every $\hat{\cV}_{\gamma_0}$ is only supported on  $\bC[\gamma_0]$ and can be written as
\bq\label{Eq:VarSummand}
 \hat{\cV}_{\gamma_0} &=& \frac{1}{2^N}\sum_{\eta \in \bZ_2^{2N}} \sum_a \rho_a\rho_{a^{\eta}} \Big{(} \oproj{a}_{\gamma_0} -  \theta_{\eta,\gamma_0} \oket{a}\obra{a^\eta}_{\gamma_0} \Big{)}.
\eq
\end{lemma}

\Proof{ This matrix is related to the variance through $\Var(f,f) = \obra{f}\hat{\cV}\oket{f}$. The definition of the variance (\ref{Variance}), for Hermitian $f \in \cM_{2^N}(\bC)$ was given by 
\bq
\Var_{\rho}(f,f)  &=& \tr{\rho f f} - \tr{\rho f}^2.
\eq
Since we are taking a full sum over all group elements we have that for any matrix $X$ defined on $\cM_d$ the following identity holds
\be
	\tr{X} \1 = \frac{1}{2^N} \sum_{\eta \in \bZ_2^{2N}} \sigma(\eta) X \sigma(\eta).
\ee
This identity is particularly helpful in finding a suitable matrix representation for $\hat{\cV}$. We can write the following: 
\bq
\1                &=& \frac{1}{2^N} \sum_\eta \sigma(\eta) \rho \sigma(\eta), \no
\tr{\rho f} \1 &=& \frac{1}{2^N} \sum_\eta \sigma(\eta) \rho  f \sigma(\eta).
\eq
Due to these identities, we can express the trace in the variance in terms of a full sum over all elements in $\bZ_2^{2N}$ and we can write that
\be
\Var_{\rho}(f,f) = \frac{1}{2^N} \sum_\eta \tr{f \rho \sigma(\eta) \rho \sigma(\eta) f } - \frac{1}{2^N} \sum_\eta  \tr{f \rho \sigma(\eta) \rho f \sigma(\eta)}.
\ee
In particular, if we define $\Var_{\rho}(f,f) = \sum_\eta \tr{f \cV^\eta(f)}$, where for each $\eta$ we have that 
\be
\cV^\eta(f) =  \frac{1}{2^N} {\Big(} \, \rho \sigma(\eta) \rho \sigma(\eta) f  - \rho \sigma(\eta) \rho f \sigma(\eta) \, {\Big)}.
\ee
If we now substitute the decomposition of the Gibbs state in terms of the projectors  $\rho = \sum_a \rho_aP(a)$,  we obtain for the matrix
\bq
\cV^\eta(f) = \frac{1}{2^N} \sum_{a,b}\rho_a \rho_b \left( P(a) \sigma(\eta) P(b) \sigma(\eta) f - P(a) \sigma(\eta) P(b) f \sigma(\eta)\right). 
\eq
We are now in the position to evaluate this matrix on the Pauli basis $\sigma(\gamma)$, in the identical fashion as we have done for the Dirichlet matrix in the previous proof by using identity (\ref{PauliMult}). We thus obtain
\bq
\cV^\eta(\sigma(\gamma)) = \frac{1}{2^{N}} \sum_a \left( \rho_a \rho_{a^\eta}  -  \rho_a \rho_{a^\eta}  \theta_{\eta,\gamma}\right)P(a) \sigma(\gamma)
\eq
Recall that we can now substitute $P(a) = 2^{-M} \sum_x e^{i\pi\avr{a , x}}\sigma(G x)$, as we have done previously to obtain the following expression purely written in the basis $\{\oket{\gamma}\}$. 
\bq \label{pruuts}
\hat{\cV}^\eta &=& \frac{1}{2^{M+N}} \sum_\gamma \sum_{a,x} \left(  \rho_a \rho_{a^\eta}  - \theta_{\alpha,\gamma}  \rho_a \rho_{a^\eta} \right)e^{i\pi \avr{x,a}} e^{i\frac{\pi}{2} {Sp}(Gx,\gamma)} \oket{ G x \oplus \gamma}\obra{\gamma}.
\eq
Note that this matrix is in its form very similar to $\hat{\cE}^\alpha$. If we define the two functions
\bq
 V^1_{\eta,\gamma}(a) = \frac{1}{2^N}\rho_a \rho_{a^\eta}  \Sp \mbox{and} \Sp  V^2_{\eta,\gamma}(a) = V^1_{\eta,\gamma}(a),
\eq
which also only depend on the syndrome $e(\eta)$ and are in fact even independent of $\gamma$ and are thus trivially constant over the cosets. We have that equation (\ref{pruuts}) is now similar to (\ref{dude}). We only need to substitute the functions $V^{1/2}_{\eta,\gamma}(a)$ for the  $E^{1/2}_{\eta,\gamma}(a)$ in Eqn. (\ref{someMatrix123}). The proof proceeds identically to the one for the Dirichlet form. The only difference is that sum is taken over all $\eta \in \bZ_2^{2N}$ in the final step, which then leads to the decomposition as stated in the lemma. \vspace{0.5cm} \qed}

As we have seen, both matrices are block diagonal in the same basis, and we can moreover write the Dirichlet matrix, as well as the variance matrix as sum of two dimensional positive matrices in the basis dual to the coset algebra. Both $\hat{\cV}$ and $\hat{\cE}$ are positive semi definite by construction and share the same kernel given by the identity matrix. Hence, the only matrix pair $(\hat{\cE}_{\gamma_0},\hat{\cV}_{\gamma_0})$ that is rank deficient corresponds to the coset that is given by $\cS$ itself. The central structural difference between the two matrices is given by the fact that for $\hat{\cE}$ the sum is only taken over $\alpha \in W_1$, i.e. single qubit Pauli matrices, whereas for $\hat{\cV}$ we sum over the full set $\eta \in \bZ_2^{2N}$. This means that there are transitions of the form $a \raw a^\eta$ which occur in $\hat{\cV}$, that are missing in $\hat{\cE}$.  

\subsection{Comparison Theorems}
Since both matrices $\hat{\cE}$ and $\hat{\cV}$ are block diagonal in the same basis, it suffices to bound the support number $\tau_{\gamma_0}$ for each subspace $\bC[\gamma_0]$ separately, since  
\be
	\tau = \max_{\gamma_0} \tau_{\gamma_0}.
\ee
To obtain the bounds on $\tau_{\gamma_0}$ we have to devise a strategy of factoring both $\hat{\cE}_{\gamma_0}$ and $\hat{\cV}_{\gamma_0}$ and embedding each into the other as discussed in lemma \ref{lem:W-bound}. It does prove convenient to consider a set of vectors that facilitate the embedding. We define for all Paulis $\alpha , \gamma_0 \in \bZ_2^{2N}$ and for all $a \in \bZ_2^{M}$ in the syndrome space the vectors 
\be
	\oket{-^\alpha_a}_{\gamma_0}= \frac{1}{\sqrt{2}}\Big{(}\oket{a}_{\gamma_0} - \theta_{\alpha,\gamma_0}\oket{a^\alpha}_{\gamma_0}\Big{)}.
\ee
These are easily obtained for every cosets and only differ by a relative phase $\theta_{\alpha,\gamma_0}$ in each coset. Moreover, the $\{\oket{-^\alpha_{a}}_{\gamma_0}\}$ do not depend on the representative of  the coset $\gamma_0$. Direct calculation reveals that $\oket{-^\alpha_{a}}_{\gamma_0} = \oket{-^\alpha_{a}}_{\gamma_1}$ if the two representatives are related by $\gamma_1 = Gx^* \oplus \gamma_0$ for some $x^* \in \bZ_2^{2M}$.

These vectors possess a convenient telescoping sum property. Given some general Pauli $\eta$ which can be expressed by a product of simpler Pauli operators $\{\alpha_i\}$, we can express the vector associated to the former Pauli as a sum of the vectors associated to the $\alpha_i$. 
 
\begin{proposition}\label{canPath}
Let $\{\alpha_i\}_{i=1,\dots,k}$ denote a set of Pauli labels $\alpha_i \in \bZ_2^{2N}$ so that the binary sum yields $\eta = \oplus_{i=1}^k \alpha_i$, then we have for all  syndromes $a \in \bZ_2^{M}$ that
\be
\oket{-_{a}^\eta}_{\gamma_0} = \sum_{s=0}^{k-1} \theta_{\; \overline{\alpha}_s,\gamma_0} \oket{-_{a^{\overline{\alpha}_s}}^{\alpha_{s+1}}}_{\gamma_0},
\ee
where $\overline{\alpha}_s = \oplus_{i=1}^s \alpha_i$, so that $\overline{\alpha}_k = \eta$.
\end{proposition}

\Proof{ We prove the claim by induction. For the trivial case  $r = 1$ where $\eta = \overline{\alpha}_r$ nothing is to prove. Let us therefore consider the induction step. Recall that by (\ref{linear:syndrome}) and (\ref{notation:syn})  we have that $e(\alpha) \oplus e(\beta) = e({\alpha \oplus \beta})$ so that ${(a^\alpha)}^\beta = a^{\alpha \oplus \beta}$. Moreover, the phases $\theta_{\alpha,\gamma_0}$ satisfy a simple multiplication rule with $\theta_{\alpha_1,\gamma_0}\theta_{\alpha_2,\gamma_0} = \theta_{\alpha_1 \oplus \alpha_2,\gamma_0}$ which follows from the bi-linearity of $\mbox{Sp}(\alpha,\gamma)_{\bZ}$. With this it is easy to show that the proposition follows from induction $r \raw r+1$ through 
\bq
\oket{-_{a}^{\overline{\alpha}_r \oplus \alpha_{r+1}}}_{\gamma_0} &=& \frac{1}{\sqrt{2}}\Big{(} \oket{a}_{\gamma_0} - \theta_{\overline{\alpha}_r,\gamma_0} \oket{a^{\overline{\alpha}_r }}_{\gamma_0} \no  
&+&  \theta_{\overline{\alpha}_r,\gamma_0} \oket{a^{\overline{\alpha}_r}}_{\gamma_0}  - \theta_{\; \overline{\alpha}_r \oplus \alpha_{r+1},\gamma_0} \oket{a^{\overline{\alpha}_r \oplus \alpha_{r+1}}}_{\gamma_0} \Big{)} \no
&=& \oket{-_a^{\overline{\alpha_r}}}_{\gamma_0}  +  \theta_{ \; \overline{\alpha_r},\gamma_0} \oket{-_{a^{\overline{\alpha_r}}}^{\alpha_{r+1}}}_{\gamma_0}\no
&=& \sum_{s=0}^{r-1} \theta_{\; \overline{\alpha}_s,\gamma_0} \oket{-_{a^{\overline{\alpha}_s}}^{\alpha_{s+1}}}_{\gamma_0} +  \theta_{ \; \overline{\alpha_r},\gamma_0} \oket{-_{a^{\overline{\alpha_r}}}^{\alpha_{r+1}}}_{\gamma_0}\no
&=& \sum_{s=0}^{r} \theta_{\; \overline{\alpha}_s,\gamma_0} \oket{-_{a^{\overline{\alpha}_s}}^{\alpha_{s+1}}}_{\gamma_0}.
\eq
\qed}

It is our goal to stay conceptionally as close as possible to the analysis of classical Markov chains \cite{sinclair1992improved}, so we can make use of the geometric picture that the classical approach provides. We therefore proceed to introduce a set of so-called canonical paths. Motivated by proposition \ref{canPath}, the form of the canonical paths for this quantum problem becomes clear. It is our goal to span a suitable linear combination of basis elements $\oket{a}_{\gamma_0}$ and $\oket{a^\eta}_{\gamma_0}$ with appropriately chosen phases by a subset of the vectors $\{ \oket{-_a^\alpha}_{\gamma_0} \}_{a \in \bZ_2^M, \eta \in \bZ_2^{2N}}$.  A canonical path then corresponds to a suitable choice of intermediate states that connects the first configuration given by $\oket{a}_{\gamma_0}$ to the final configuration $\oket{a^\eta}_{\gamma_0}$.\\

It is important to differentiate between the different kinds of paths here. The small latin letters $ a \in \bZ_2^M$, label the syndromes that stem from the generators in $\cG$, whereas the $\gamma, \eta \in \bZ_2^{2N}$ label the Pauli operators that give rise to particular syndromes $e(\gamma),e(\eta)$. Since the phases $\theta_{\eta,\gamma_0} = \pm 1$ in proposition \ref{canPath} are needed we need to keep track of both the syndromes, as well as the corresponding Pauli operator that generates them. We will therefore distinguish between simple Pauli paths, which build up a particular Pauli operator by applying single qubit Pauli operators and Pauli operators, which are dressed with syndrome values. 

\begin{definition}\label{def:cannonicalPaths}
We introduce new labels $(a,\eta)$, where  $a \in \bZ_2^M$ denotes a syndrome of the code $G$ and  $\eta \in \bZ^{2N}_2$ a Pauli matrix. We define the following:

\begin{enumerate}
\item A {\bf Pauli path} $\overline{\eta}$ is a sequence of single qubit Pauli operators labeled by $\{\alpha_i\}_{i = 1 \ldots T }  \subset W_1$, so that $\eta = \oplus_{i=1}^T \alpha_i$. 
We denote by $\overline{\eta}_t = \oplus_{i=1}^t \alpha_i$ the partially constructed Pauli operator at step $t \in \{0,1,\ldots,T\}$ of the path. We define $\eta_0 = (0,0)^N$.

\item A {\bf canonical path}, or dressed Pauli path, from $(a,0) \raw (a^\eta,\eta)$ is constructed for every syndrome $a$ in $G$ and any Pauli $\eta$ from a Pauli path $\overline{\eta}$ 
as the sequence of pairs  
\be
	\hat{\eta}_a = \left[(a,0),(a^{\overline{\eta}_1},\overline{\eta}_1),\ldots,(a^{\overline{\eta}_{T-1}},{\overline{\eta}_{T-1}}), (a^\eta,\eta)\right].
\ee
The length of the canonical path is defined by $|\hat{\eta}_a| = T$. The set of canonical paths that uniquely connects all paired labels $(a,0) \raw (a^\eta,\eta)$ is denoted by $\Gamma$. 

\item Furthermore, a  subsequent pair of labels $\hat{\xi} = [(a^\xi, \xi),(a^{\xi \oplus \alpha}, {\xi \oplus \alpha})]$, which only differs by a single qubit Pauli $\alpha \in W_1$  is called an {\bf edge}. We denote by $\Gamma(\hat{\xi}) \subset \Gamma$, the subset of canonical paths $\hat{\eta}_a$ that contain the edge $\hat{\xi}$.
\end{enumerate}
\end{definition}

Since, every Pauli matrix $\sigma(\eta)$ can be decomposed into at most $N$ single qubit Pauli's the different $\alpha_i$ can be determined easily. However, what is not directly obvious is the order by which the single qubit Pauli's are applied. It turns out in fact, that this order matters in the derivation of good bounds as we will see in the subsequent section. This particular order strongly depends on the particular code that is investigated in order to obtain the best possible bound admissible by our aproach. With these paths, we can now state the upper bound on the support number $\tau$.

\begin{theorem}\label{no-syn:support_bound}
The support number $\tau$ for the matrix pair $(\hat{\cV},\hat{\cE})$ with a choice of canonical paths $\Gamma$  is bounded by
\be
\tau \leq \max_{(a,\mu)} \sum_{\hat{\xi} \in \hat{\mu}_a} \frac{4}{2^N h(\omega^\alpha(a^\xi))\rho_{a^\xi}} \sum_{\hat{\eta}_a \in \Gamma(\hat{\xi})} \rho_a \rho_{a^\eta}.
\ee
The maximum is take over all syndrome - Pauli labels $(a,\mu)$ and we denote by $\hat{\xi} \in \hat{\mu}_a$ the sum over all edges $\hat{\xi} = [(a^\xi, \xi),(a^{\xi \oplus \alpha}, {\xi \oplus \alpha})]$ that are crossed in the canonical path $\hat{\mu}_a$.
\end{theorem}

\vspace{0.5cm}
\Proof{ Recall that $\tau = \max_{[\gamma_0]} \tau_{\gamma_0}$, due to the decomposition $\hat{\cE} = \oplus_{[\gamma_0]} \hat{\cE}_{\gamma_0}$ and $\hat{\cV} = \oplus_{[\gamma_0]} \hat{\cV}_{\gamma_0}$. We therefore only need to consider the support number $\tau_{\gamma_0}$ for every individual coset of the pair $\hat{\cE}_{\gamma_0},\hat{\cV}_{\gamma_0}$.  The matrices $\hat{\cV}_{\gamma_0}$ and $\hat{\cE}_{\gamma_0}$ can be brought into a particularly simple form which bears some resemblance to that of a graph Laplacian \cite{chung1997spectral}. The form is, however, different in that both matrices have positive as well as negative off diagonals which stem from the phases $\theta_{\eta,\gamma_0} = \pm 1$ in both (\ref{Eq:DirichletSummand}) and (\ref{Eq:VarSummand}). Nevertheless the matrices can be related to a sum of  rank one  projectors. Consider first
\bq
\hat{\cV}_{\gamma_0}  &=& \frac{1}{2^N} \sum_{\eta} \sum_a \rho_a\rho_{a^{\eta}} \Big{(} \oproj{a}_{\gamma_0} -  \theta_{\eta,\gamma_0} \oket{a}\obra{a^\eta}_{\gamma_0} \Big{)} \no
&=& \frac{1}{2^N} \sum_\eta \sum_a \rho_a\rho_{a^{\eta}} \oproj{-_a^\eta}_{\gamma_0},
\eq
which follows by direct calculation.\\

The $\hat{\cE}_{\gamma_0}$ can only be brought into this form for particular cosets, which are related to Pauli operators $\gamma_{0}$ that have a vanishing syndrome.  These Paulis correspond to operators in the center ${\cal C}_{\cP_N}(\cS)$. For these cosets we have  $e(\gamma_0) = 0 $ so that $\omega^\alpha(a) = \omega^\alpha(a^{\gamma_0})$ and the matrices in (\ref{Eq:DirichletSummand}) simplify to
\bq\label{someExpr}
\hat{\cE}_{\gamma_0} =  \sum_{\alpha \in W_1} \sum_a  h^\alpha(\omega^\alpha(a)) \rho_a \oproj{-_a^\alpha}_{\gamma_0}.
\eq
This is not the case in general, however. When we consider cosets for which $e(\gamma_0) \neq 0$, we naturally have that there exist pairs of Bohr frequencies for which $\omega^\alpha(a) \neq \omega^\alpha(a^{\gamma_0})$ so that $h^\alpha_{aa^{\gamma_0}} =0$. However, it is still possible to bound these cosets at the expense of a factor of four by the expression (\ref{someExpr}). Consider the basis $\oket{a}_{\gamma_0},\oket{a^\alpha}_{\gamma_0}$ so that we can express the symmetrization of eqn. (\ref{Eq:DirichletSummand}),
\bq
\half\left(\hat{\cE}^\alpha_{\gamma_0}(a) + \hat{\cE}^\alpha_{\gamma_0}(a^\alpha) \right)
 = \half \left(  \begin{array}{cc} \half\left( h^\alpha_{aa} + h^\alpha_{a^\gamma a^{\gamma_0}} \right) \rho_a & -h^\alpha_{aa^{\gamma_0}}\rho_a \theta_{\alpha,{\gamma_0}} \vspace{0.2cm}\\ -h^\alpha_{aa^{\gamma_0}}\rho_a \theta_{\alpha,{\gamma_0}} & \half\left( h^\alpha_{a^\alpha a^\alpha} + h^\alpha_{a^{\alpha{\gamma_0}} a^{ \alpha{\gamma_0}}} \right) \rho_{a^\alpha} \end{array} \right)
\eq 
as a simple two dimensional matrix. In the particular case, where $\omega^\alpha(a) = \omega^\alpha(a^{\gamma_0})$, we have that $h_{a,a^{\gamma_0}} = h_{a,a} = h_{a^{\gamma_0},a^{\gamma_0}}  =  h^\alpha(\omega^\alpha(a))$ and the syndrome $e(\gamma_0)$ does not contribute so that we have again that $\half\left(\hat{\cE}^\alpha_{\gamma_0}(a) + \hat{\cE}^\alpha_{\gamma_0}(a^\alpha) \right) = h^\alpha(\omega^\alpha(a)) \rho_a \oproj{-_a^\alpha}_{\gamma_0}$. When, however, $\omega^\alpha(a) \neq \omega^\alpha(a^{\gamma_0})$, we can find the bound   
\bq
\half\left(\hat{\cE}^\alpha_{\gamma_0}(a) + \hat{\cE}^\alpha_{\gamma_0}(a^\alpha) \right) &=&  \frac{1}{4}\left( h^\alpha_{aa} + h^\alpha_{a^\gamma a^\gamma} \right) \rho_a  \oproj{a}_{\gamma_0} \no 
&+&   \frac{1}{4}\left( h^\alpha_{a^\alpha a^\alpha} + h^\alpha_{a^{\alpha {\gamma_0}} a^{ \alpha {\gamma_0}}} \right) \rho_{a^\alpha} \oproj{a^\alpha}_{\gamma_0} \no
&\geq&   \frac{1}{4} h^\alpha_{aa} \rho_a  \oproj{a}_{\gamma_0} +  \frac{1}{4} h^\alpha_{a^\alpha a^\alpha}  \rho_{a^\alpha} \oproj{a^\alpha}_{\gamma_0} \no
&\geq&   \frac{1}{4} h^\alpha_{aa} \rho_a  \oproj{-_{a}^\alpha}_{\gamma_0}.
\eq
The first inequality is obtained by dropping the positive numbers $h^\alpha_{a^\gamma a^\gamma}$ and $ h^\alpha_{a^{\alpha \gamma} a^{ \alpha \gamma}}$. The final inequality follows from the KMS condition since $h^\alpha_{aa} \rho_a = h^\alpha_{a^\alpha a^\alpha}  \rho_{a^\alpha}$ and the trivial bound $\oproj{a}_{\gamma_0} + \oproj{a^\alpha}_{\gamma_0} \geq \oproj{-_a^\alpha}_{\gamma_0}$.\\ 

Thus, we have the following semi-definite inequality for the Dirichlet matrix
\be
	\hat{\cE}_{\gamma_0} \geq  {\hat{\cE}}'_{\gamma_0}  \equiv  \sum_{\alpha \in W_1} \sum_{a} \frac{1}{4} h^\alpha(\omega^\alpha(a))\rho_a \oproj{-_{a}^\alpha}_{\gamma_0}.
\ee
It turns out that due to the very similar form of the matrices, it is in fact simpler to bound the constant ${\tau}'_{\gamma_0}$ for the matrix pair ${\hat{\cE}}'_{\gamma_0}$ and $\hat{\cV}_{\gamma_0}$. This bound is a natural upper bound to ${\tau}'_{\gamma_0} \geq {\tau}_{\gamma_0}$, since we have that 
\bq
0  &\leq& {\tau}'_{\gamma_0}{\hat{\cE}}'_{\gamma_0}  - \hat{\cV}_{\gamma_0} \no  
       &=&   {\tau}'_{\gamma_0}\hat{\cE}_{\gamma_0}    - \hat{\cV}_{\gamma_0}  -{\tau}'_{\gamma_0}\left({\hat{\cE}}_{\gamma_0} - {\hat{\cE}}'_{\gamma_0}\right)
        \leq  {\tau}'_{\gamma_0}\hat{\cE}_{\gamma_0}     - \hat{\cV}_{\gamma_0}.
\eq
The last inequality follows from the previously derived fact that $\hat{\cE}_{\gamma_0} - {\hat{\cE}}'_{\gamma_0} \geq 0$. We will proceed to bound only  ${\tau}'_{\gamma_0}$ for the matrix pair ${\hat{\cE}}'_{\gamma_0}$ and $\hat{\cV}_{\gamma_0}$, since ${\tau}'_{\gamma_0}$ gives rise to a valid lower bound to the spectral gap.\\

The matrices $\hat{\cE}'_{\gamma_0}$ and $\hat{\cV}_{\gamma_0}$ are now in an almost identical form. The central structural difference is that the sum in $\hat{\cE}'_{\gamma_0}$ is taken only over all single qubit Pauli operators $\alpha \in W_1$, whereas for  $\hat{\cV}_{\gamma_0}$ we need to sum over the full algebra. It is now tempting to identify an edge with the transition of syndromes $a \raw a^\alpha$ for every vector $\oket{-_a^\alpha}_{\gamma_0}$. However, recall that it is necessary to keep track of the phases $\theta_{\overline{\eta}_t,\gamma_0} = \pm 1 $ in proposition \ref{canPath}. To this end we seek to construct a factorization that allows to distinguish the different phases. We introduce an additional sum over the full Pauli algebra $\varphi \in \bZ_2^{2N}$, so that 
\bq
\hat{\cE}'_{\gamma_0} &=& \frac{1}{4^N} \sum_a \hat{\cE}'_{\gamma_0}(a) \Sp \mbox{where,} \Sp \hat{\cE}'_{\gamma_0}(a) = \sum_{\varphi \in \bZ_2^{2N}} \sum_{\alpha \in W_1} \frac{1}{4} h^\alpha(\omega^\alpha(a^\varphi))\rho_{a^\varphi} \oproj{-_{a^\varphi}^\alpha}_{\gamma_0}, \no
\hat{\cV}_{\gamma_0} &=& \frac{1}{4^N} \sum_a \hat{\cV}_{\gamma_0}(a) \Sp \mbox{where,} \Sp \hat{\cV}_{\gamma_0}(a) = \sum_{\varphi , \eta \in \bZ_2^{2N}} \frac{1}{2^N} \rho_{a^\varphi} \rho_{a^{\varphi \eta}} \oproj{-_{a^\varphi}^\eta}_{\gamma_0}.
\eq
With this, we consider the bound $\tau_{\gamma_0} \leq \max_{a} \tau'_{\gamma_0}(a)$, where $ \tau'_{\gamma_0}(a) \hat{\cE}'_{\gamma_0}(a) - \hat{\cV}_{\gamma_0}(a) \geq 0$. To find upper bounds to $\tau_{\gamma_0}(a)$, we construct a factorization as stated in lemma \ref{lem:W-bound} and apply the norm bound in lemma \ref{normBND}. 

We introduce a new orthonormal auxiliary basis spanned by $\{\oket{\varphi,\mu}\}$ with  $ \obraket{\varphi,\mu}{\beta, \kappa} = \delta_{\varphi,\beta} \delta_{\mu,\kappa}$, for every pair $\varphi, \mu \in \bZ_2^{2N}$.

\begin{lemma}\label{no-syn:embedding}
For  ${\hat{\cE}}'_{\gamma}(a)$ and $\hat{\cV}_{\gamma}(a)$  as above, we can find a factorization into the triple $[A_{\gamma}(a), B_{\gamma}(a) , W_{\gamma}(a)]$, subject to a chosen set of canonical paths $\Gamma = \{\hat{\eta}_a\}$. The matrices are given by
\bq
A_{\gamma_0}(a)   &=&  \sum_{\varphi \in \bZ^{2N}_2} \sum_{\alpha \in W_1}  \sqrt{\frac{1}{4} h(\omega^\alpha(a^\varphi))\rho_{a^\varphi}} \; \oket{-^{\alpha}_{a^\varphi}}_{\gamma_0}\obra{\varphi,\varphi \oplus \alpha} \\
B_{\gamma_0}(a)   &=&  \sum_{\varphi,\eta \in \bZ^{2N}_2} \sqrt{\frac{1}{2^N}\rho_{a^{\varphi \eta}} \rho_{a^\varphi}} \; \oket{-^{\eta}_{a^\varphi}}_{\gamma_0}\obra{\varphi, \varphi \oplus \eta} \no
W_{\gamma_0}(a)  &=&  \sum_{\varphi,\eta \in \bZ^{2N}_2} \sum_{t=0}^{|\hat{\eta}_a|-1} \sqrt{\frac{4 \rho_{a^\varphi} \rho_{a^{\varphi \eta}}}{2^N h(\omega^{\alpha_{t+1}}(a^{\varphi\overline{\eta}_t}))\rho_{a^{\varphi \overline{\eta}_t}}}} \; \theta_{\overline{\eta}_t,\gamma_0} \;\oket{\varphi \oplus \overline{\eta}_t ,\varphi \oplus \overline{\eta}_{t+1}}\obra{\varphi,\varphi \oplus \eta },\nonumber
\eq
where of course $\overline{\eta}_{t+1} = \overline{\eta}_{t} \oplus \alpha_{t+1}$.
\end{lemma}

\proof{ We immediately have by direct computation that  $A_{\gamma_0}(a)A_{\gamma_0}(a)^\dag = {\hat{\cE}}'_{\gamma_0}(a)$ and $B_{\gamma_0}(a)B_{\gamma_0}(a)^\dag = \hat{\cV}_{\gamma_0}(a)$. Moreover, we have that
\bq
A_{\gamma_0}(a)W_{\gamma_0}(a) &=&	\sum_{\varphi,\eta \in \bZ^{2N}_2} \sum_{t=0}^{|\hat{\eta}_a|-1} \sqrt{\frac{1}{2^N}\rho_{a^\varphi} \rho_{a^{\varphi \eta}}} \; \theta_{\overline{\eta}_t,\gamma_0} \; \oket{-_{a^{\varphi \overline{\eta}_t}}^{\alpha_{t+1}}}_{\gamma_0}\obra{\varphi, \varphi \oplus \eta} \no
					       &=& 	\sum_{\varphi,\eta \in \bZ^{2N}_2} \sqrt{\frac{1}{2^N} \rho_{a^\varphi} \rho_{a^{\varphi \eta}}} \left( \sum_{t=0}^{|\hat{\eta}_a|-1} \; \theta_{\overline{\eta}_t,\gamma_0} \; \oket{-_{a^{\varphi \overline{\eta}_t}}^{\alpha_{t+1}}}_{\gamma_0} \right)\obra{\varphi, \varphi \oplus \eta} \no
					       &=&  \sum_{\varphi,\eta \in \bZ^{2N}_2} \sqrt{\frac{1}{2^N} \rho_{a^\varphi} \rho_{a^{\varphi \eta}}}\; \oket{-^{\eta}_{a^\varphi}}_{\gamma_0}\obra{\varphi ,\varphi \oplus \eta} = B_{\gamma_0}(a).
\eq
The final equality is due to the decomposition into canonical paths and proposition \ref{canPath}. \qed}

Let us now apply the norm bound of lemma \ref{normBND} to  $W_{\gamma_0}(a)$ as given in lemma \ref{no-syn:embedding}, in order to obtain an upper bound to $\tau_{\gamma_0}(a)$ To do so we must first compute the norm of the row vectors of $W_{\gamma_0}(a)$. That is we need to fix the transition $ (\varphi \oplus \xi, \varphi \oplus \xi \oplus \alpha )$, for which we can read of directly
\be
\ket{w_{ (\varphi \oplus \xi, \varphi \oplus \xi \oplus \alpha )}} = \frac{2 \theta_{\xi,\gamma_0}}{\sqrt{2^N h(\omega^{\alpha}(a^{\varphi \xi})) \rho_{a^{\varphi \xi}} }} \sum_{(\varphi \raw \varphi \oplus \eta) \ni \hat{\xi} } ' \;  \sqrt{\rho_{a^\varphi} \rho_{a^{\varphi \eta}}} \; \oket{\varphi,\varphi \oplus \eta}.
\ee
The constrained sum over pairs $(\varphi \raw \varphi \oplus \eta) \ni \hat{\xi}$ , is taken to read, that there exists a Pauli path $\overline{\eta}_t$, which transforms $\varphi$ into $\varphi \oplus \eta$ so that the list of Pauli operators that are traversed contains the two subsequent Pauli configurations $\varphi \oplus \xi, \varphi \oplus \xi \oplus \alpha$. It is now easy to compute the norm bound simply by squaring the individual summands. 
\be
\|\ket{w_{ (\varphi \oplus \xi, \varphi \oplus \xi \oplus \alpha )}}\|_2^2 = \frac{4}{2^N h(\omega^{\alpha}(a^{\varphi \xi})) \rho_{a^{\varphi \xi}}}\sum_{(\varphi \raw \varphi \oplus \eta) \ni \hat{\xi} } '  \; \rho_{a^\varphi} \rho_{a^{\varphi \eta}}.
\ee
If we apply the norm bound in lemma \ref{normBND}, we have that the condition $W_{k,m} \neq 0$ in the sum means that we have to sum the norms $\|\ket{w_{ (\varphi \oplus \xi, \varphi \oplus \xi \oplus \alpha )}}\|_2^2$ over all transitions that are crossed when transitioning from a initial Pauli $\varphi$ to $\varphi \oplus \mu$, in the Pauli path $\overline{\mu}$ that maximizes this expression. It is of course also possible to relabel these sums in terms of the edges and dressed Pauli Paths as defined in definition \ref{def:cannonicalPaths}. So that we write
\bq
\|W_{\gamma_0}(a)\|^2 &\leq& \max_{(\varphi ,\varphi \oplus \mu)} \sum_{\hat{\xi} \in \hat{\mu}_a} \|\ket{w_{ (\varphi \oplus \xi, \varphi \oplus \xi \oplus \alpha )}}\|_2^2 \no
&=&\max_{(\varphi ,\varphi \oplus \mu)} \sum_{\hat{\xi} \in \hat{\mu}_{a^\varphi}} \frac{4 }{2^N h(\omega^\alpha(a^{\varphi \xi}))\rho_a^{\varphi \xi}} \sum_{\hat{\eta}_{a^\varphi} \in \Gamma(\hat{\xi})} \rho_{a^{\varphi}} \rho_{a^{\varphi \eta}}.
\eq
Now, we furthermore have that $\tau'_{\gamma_0} \leq \max_a \|W_{\gamma_0}(a)\|^2$ and we observe that the only dependence on the initial Pauli matrix $\varphi$ is through the syndrome $a^\varphi$, we can drop the dependence and consider any path starting from the identity $\varphi = 0$. Hence, we just absorb the maximization over $a$ and consider now only dressed Pauli paths. We observe, that the bound $\tau_{\gamma_0}$ does not depend on the coset, so that for any coset $[\gamma_0]$ we have that $\tau \leq \tau_{\gamma_0}$ so that we are left with the final bound as stated in the theorem. \qed}

\section{The spectral gap and the energy barrier}
\label{sec:bound}

In Theorem \ref{no-syn:support_bound} we have worked towards finding a bound on $\tau$ that is formally similar to the bound obtained from the canonical paths lemma for classical Markov processes. This allows us to follow an approach first pioneered by Jerrum and Sinclair \cite{sinclair1989approximate} to evaluate this lower bound. In the theorem, we have left the particular choice for the set of canonical paths $\Gamma = \{\hat{\eta}_a\}$, c.f. definiton \ref{def:cannonicalPaths}, we want to work with open. It turns out that the particular choice of paths $\hat{\eta}_a$ strongly depends on the stabilizer Hamiltonian $H$ we try to investigate. The wrong choice of paths can lead to an exponentially worse lower bound when compared to reasonable choice, c.f.  section \ref{example}. \\

A canonical path connects an initial syndrome $a$ to a syndrome $a^\eta = a \oplus e(\eta)$, by constructing a Pauli operator $\sigma(\eta)$ from single qubit Pauli matrices $\sigma(\alpha)$, with $\alpha \in W_1$. Hence, we need to agree on a path for every syndrome $a$ and any Pauli $\eta$. As we will see, the choice of the decomposition of $\eta$ into single qubit Paulis  $W_1$ does not depend on the initial syndrome $a$ and we will use the same decomposition of $\eta$ for different initial syndromes $a$. So the construction reduces to finding good Pauli paths that connect the identity $\overline{\eta}_0 = (0,0)^N$ to the final Pauli $\overline{\eta}_{|\hat{\eta}_a|} = \eta$. That is, we only need to specify for every $\eta \in \bZ_2^{2N}$ a specific order in which the single qubit Pauli's are applied. An important constraint in the construction of any path $\hat{\eta}_a$ is that this path is free from loops, i.e. it does not contain the same edge $\hat{\xi}$ twice. Note that we do not refer to geometric loops in the partially constructed Pauli operators $\overline{\eta}_t$, but to loops on the Cayley graph associated with $(\bZ_2^{2N},W_1)$. It is always possible to find paths that are free of loops by following the most trivial decomposition of $N$-qubit Paulis into their single qubit components. Moreover, as will become clear it is obvious that such loops do not improve the bounds. 

We are now in the position to define the central quantity that determines the lower bound to the spectral gap for every commuting Pauli Hamiltonian. The generalized energy barrier is very similar to the energy barrier for logical operators as defined in \cite{bravyi2009no}. However, in our definition we do not assume that we consider a stabilizer code with logical operators. Any commuting Pauli Hamiltonian can be analyzed this way.

\begin{definition}\label{def:GenEng}
Given a commuting Pauli Hamiltonian $H$ as  in eqn. (\ref{CommPauli}), with generator set $\cG$, that can be used to define the excitations $e_k(\eta)$ for some $\eta \in \bZ_2^{2N}$, c.f. eqn (\ref{linear:syndrome}), we define:

\begin{enumerate}
\item For any $\eta$ with a Pauli path $\overline{\eta}_t = \oplus_{s=0}^t \alpha_s$ the {\bf energy cost} of the Pauli $\eta$ as 
\be\label{energy:bnd}
 	\overline{\epsilon}(\eta) =  \min_{\{\overline{\eta}\}} \max_{t} \sum_{k=1}^M 2 |J_k|e_k(\overline{\eta}_t) \overline{e}_k(\eta).
\ee
Here $\overline{e}_k = e_k \oplus 1$ denotes the conjugation of the bit value. In this sum $e_k$ and $\overline{e}_k$ are interpreted as integers. The minimum is taken over all possible choices of Pauli paths for $\eta$.

\item Furthermore, we define the {\bf generalized energy barrier} as the maximum over all Pauli's $\eta$ of the energy cost
\be \label{GenEng}
\overline{\epsilon}  = \max_{\eta \in \bZ^{2N}_2} \overline{\epsilon}(\eta).
\ee
\end{enumerate}
\end{definition}

The definition of the generalized energy barrier differs from the energy barrier of a logical Pauli operator as given in \cite{bravyi2009no} in two aspects. First the energy cost differs by the factors $\overline{e}_k(\eta)$ in the summation (\ref{energy:bnd}). The essential  effect of these factors is to remove any contribution to the barrier that originate from the final Pauli operator $\eta$ itself. Therefore the only summands that contribute to $\overline{\epsilon}(\eta)$ come from violations of generators $g_i$ which are not already violated by $\eta$  by itself. That is we only care about the intermediate energy configurations in the construction of this Pauli. The energy cost $\overline{\epsilon}(\eta)$  is therefore a meaningful quantity even if $\eta$ is not a logical operator. Furthermore, the generalized energy barrier $\overline{\epsilon}$ contains a maximum over all Pauli operators, as opposed to a minimum over logical operators. This may be seen as the origin of why the generalized energy barrier will tend to pick up on the largest energy barrier of all the logical operator, as we will see in the example section \ref{example}.\\

In essence it is this energy barrier that determines the choice of the canonical paths $\Gamma = \{\hat{\eta}_a\}$ through the minimum over all Pauli paths $\{\overline{\eta}\}$ in eqn \ref{energy:bnd}. That is, we choose $\Gamma$ such that  for every $\eta$ the energy cost is minimized. Since we have to do so for every Pauli, this already fixes $\Gamma$. For practical purposes, however, we need to provide a concrete instruction for generating the Pauli path $\overline{\eta}$ in order to evaluate $\overline{\epsilon}$. Any sub optimal choice, will only lead to an upper bound $\overline{\epsilon}' \geq \overline{\epsilon}$ to the true generalized energy barrier. This will turn out to only lower our bound on the spectral gap.\\ 

\paragraph{Convenient interpretation of the  generalized energy barrier:} 
A concrete example for the evaluation is provided in section \ref{example}. In general, we can interpret the barrier as follows:  Suppose, we are given the set of commuting Pauli operators $\cG = \{g_i\}_{i=1,\ldots,M}$ that define the Hamiltonian $H$. If we want to evaluate $\overline{\epsilon}(\eta)$ for some particular $\eta$ we consider a reduced subset of generators 
\be
	\cG_\eta = \Big{\{} g \in \cG \; \Big{|} \; [g,\sigma(\eta)] = 0 \Big{\}},
\ee
which is obtained by removing all generators $g_i$ from the generating set that anti commute with the Pauli operator $\sigma(\eta)$. If the original set generated a stabilizer group $\cS = \avr{\cG}$, we can now consider the reduced subgroup $\cS_\eta = \avr{\cG_\eta}$, for which  $\sigma(\eta)$ behaves like a logical operator, i.e. by construction $\sigma(\eta) \in {\cal C}(\cS_\eta) \backslash \cS_\eta$. The energy cost $\overline{\epsilon}(\eta)$ can then be interpreted as the conventional energy barrier  \cite{bravyi2009no} of the logical operator $\sigma(\eta)$ of the new code $\cS_\eta$. This of course immediately implies, that if $\eta$ was a logical operator for the original stabilizer group $\cS$, then $\overline{\epsilon}(\eta)$ is just the conventional energy barrier for this particular logical operator. The conventional energy barrier for all the logical operators therefore always constitutes a lower bound to the generalized energy barrier. Furthermore, when any local defect can be grown into a logical operator of a stabilizer code $\cS$ by applying single qubit Pauli operators and in turn any Pauli operator can be decomposed into a product of clusters of such excitations, $\overline{\epsilon}$ corresponds to the largest energy barrier of any of the canonical logical operators \cite{yoshida2010framework}. \\

We proceed to give a bound on $\tau$ as stated in theorem \ref{no-syn:support_bound}. By applying a trick first pioneered by Jerrum and Sinclair \cite{sinclair1989approximate}, we will see that the generalized energy barrier from definition \ref{def:GenEng} will assume the central role. Let us consider a single edge $\hat{\xi} = [(a^\xi, \xi),(a^{\xi \oplus \alpha} ,  \xi \oplus \alpha) ]$,  c.f. definition \ref{def:cannonicalPaths}. We can define an injective map  from the set of canonical paths  $\Gamma(\xi)$ that contain this edge into the full Pauli group.
\be\label{SinclairMap}
 \Phi_\xi : \Gamma(\xi) \raw \bZ^{2N}_2.
\ee 
The action of the map is defined as follows. For a path $\hat{\eta}_a \in \Gamma(\xi)$, the resulting Pauli operator $\Phi_\xi(\hat{\eta}_a)$ is simply given by
\bq\label{injective-map}
  \Phi_\xi(\hat{\eta}_a)   =\eta \oplus \xi
\eq
One can see that this map is injective by constructing it's inverse on the set $\Gamma(\xi)$.  To this end it is important that every edge occurs only once in a canonical Path in order to avoid double counting the same path. For every Pauli operator in the image of the map $ \varphi = \Phi_\xi(\hat{\eta}_a)$ and the information about the transition $\hat{\xi}$, we can construct the tuple $(a,\eta)$ of the canonical path $\hat{\eta}_a$. Given $(\hat{\xi},\varphi)$ find the Pauli $\eta$ simply by $ \eta = \varphi \oplus \xi $. Moreover, given $\hat{\xi}$ we can immediately reconstruct the syndrome $a$ from $a = a^\xi \oplus e(\xi)$. Every canonical path in $\Gamma$ is uniquely determined through the initial syndrome $a$ and the final Pauli $\eta$. This uniquely identifies the canonical path $\hat{\eta}_a \in \Gamma(\xi)$. 

\begin{proposition} For all pairs $(a,\eta)$ and edges $\hat{\xi} = [(a^\xi,\xi),(a^{\xi \oplus \alpha},\xi \oplus \alpha)]$ contained in the canonical path $\hat{\eta}_a$, the following inequality holds:
\be\label{important-bound}
	 \rho_{a^\xi} \rho_{a^{\Phi_\xi(\hat{\eta}_a)}}   \geq   e^{-\beta 2 \overline{\epsilon}} \rho_a \rho_{a^\eta}.
\ee
\end{proposition}

\proof{ For some edge $\hat{\xi}$, we are able to relate the syndromes $a$ in the initial configuration and $a^\xi$ at the edge $\hat{\xi}$ by $a = a^\xi \oplus e(\xi)$.  Moreover, using the particular form of the map $\Phi_\xi$, from eqn. (\ref{injective-map}), we can relate the final Pauli operator $\eta$ of the path and the Pauli operator at the edge $\xi$ through $\eta \oplus \xi = \Phi_\xi(\hat{\eta}_a)$. Recall that the Gibbs weight is simply given by $\rho_a = Z^{-1}\exp(-\beta \epsilon_a)$ as was derived in Eqn. (\ref{Gibbs-rep}). For the inequality (\ref{important-bound}) to hold we need to find some constant , say $m$, so that for all paths $\hat{\eta}_a \in \Gamma$ and traversed edges $\hat{\xi}$ we have that $exp(-\beta( \epsilon_{a^\xi} + \epsilon_{a^{\Phi_\xi(\hat{\eta}_a)}}))  e^{\beta 2 m} \geq  exp(-\beta (\epsilon_a + \epsilon_{a^\eta}))$. Comparing the exponents, we need to find a constant $m$ such that
\be
	2m \geq \epsilon_{a^{\Phi_\xi(\hat{\eta}_a)}} + \epsilon_{a^\xi} - \epsilon_a - \epsilon_{a^\eta}.
\ee
For any edge $\hat{\xi}$ and any canonical path $\hat{\eta}_a$ that traverse it we can evaluate $\Phi_\xi(\hat{\eta}_a)$ as discussed in the previous paragraph so that we can write  
\bq
&&\epsilon_{a^{\eta \oplus \xi}} + \epsilon_{a^\xi} - \epsilon_a - \epsilon_{a^\eta}\no  
&&= \sum_{k=1}^M J_k (-1)^{a_k}\left( 1 + (-1)^{e_k(\eta)} - (-1)^{e_k(\eta \oplus \xi)} (-1)^{e_k(\xi)} \right)\no
&&= 2 \sum_{k=1}^M 2J_k (-1)^{a_k} \half\left(1 - (-1)^{e_k(\xi)}\right) \half\left(1 + (-1)^{e_k(\eta)}\right).
\eq
We need to look for the assignment that maximizes this expression. We need to choose the $a_k$ such that every summand is positive. Furthermore we simply have, that by interpreting the syndromes as $0,1$ valued integers that  
\be
\half\left(1 - (-1)^{e_k(\xi)}\right) = e_k(\xi) \Sp \mbox{and} \Sp  \half\left(1 + (-1)^{e_k(\eta)}\right) =  \overline{e}_k(\eta).
\ee 
Since $\hat{\eta}_a$ traverse the edge $\hat{\xi}$, we have that $\xi = \overline{\eta}_t$ for some value $t$. Also, recall that this bound holds for any choice of Pauli paths $\{\overline{\eta}\}$ that traverse the edge $\eta_t$. Hence minimizing over all possible choices of paths will yield an improved constant: $ \overline{\epsilon}(\eta) = \min_{\{ \overline{\eta}\} } \max_t \sum_{k=1}^M 2 |J_k| e_{k}(\overline{\eta}_t)\overline{e}_k(\eta)$. Moreover, we require this to hold for all edges and all Pauli operators. We therefore need to choose $m$ as $ m = \overline{\epsilon} =  \max_\eta \overline{\epsilon}(\eta)$, which results in the bound as stated in the proposition. \vspace{0.5cm} \qed}

This proposition provides the important inequality that relates the generalized energy barrier to the bound from theorem \ref{no-syn:support_bound} and thus in turn to the spectral gap of the Davies generator.  We will first consider an evaluation of the bound in theorem \ref{no-syn:support_bound} that is very close to the proof of Ref. \cite{sinclair1989approximate} and its exposition in Ref. \cite{Martinelli}. 

\begin{theorem}\label{Gen-bound}
For any commuting Pauli Hamiltonian $H$,  eqn. (\ref{CommPauli}), the spectral gap $\lambda$ of the Davies generator $\cL_\beta$, c.f. eqn (\ref{Davies}), with weight one Pauli couplings $W_1$
is bounded by
\be\label{Gen-bound-eqn}
	\lambda \geq  \frac{h^*}{4 \eta^*} \exp(- 2 \beta \; \overline{\epsilon} ),
\ee
where $\overline{\epsilon}$, denotes the generalized energy barrier defined in (\ref{GenEng}). Here $\eta^*$ denotes the length of the largest path in Pauli space and $h^* = \min_{\omega^\alpha(a)}h^\alpha(\omega^\alpha(a))$ is the smallest transition rate (\ref{KMS}).
\end{theorem}

\Proof{ We proceed to evaluate the bound in theorem \ref{no-syn:support_bound} for $\tau$. Observe that we can bound the first sum in the theorem as follows:
\be\label{max-max_bnd}
\tau \leq \left( \max_{(a,\mu)} \sum_{\hat{\xi} \in \hat{\mu}_a}\right) \max_{\hat{\xi}} \frac{4}{2^N h(\omega^\alpha(a^\xi))\rho_{a^\xi}} \sum_{\hat{\eta}_a \in \Gamma(\hat{\xi})} \rho_a \rho_{a^\eta}	
\ee
by choosing the largest possible edge $\hat{\xi}$ for any canonical path. Moreover, we trivially have that $ \max_{(a,\mu)} \sum_{\hat{\xi} \in \hat{\mu}_a}  \leq \eta^*$, where $\eta^*$ denotes the length of the largest canonical path, so that 
\be
\tau \leq  \max_{\hat{\xi}} \frac{4 \eta^*}{2^N h(\omega^\alpha(a^\xi))\rho_{a^\xi}} \sum_{\hat{\eta}_a \in \Gamma(\hat{\xi})} \rho_a \rho_{a^\eta}.	
\ee
If we define $h^*$ as given in the theorem and use eqn. (\ref{important-bound}), so that we have for all paths that use this edge that $ e^{ \beta 2 \overline{\epsilon}}{h^*}^{-1} \rho_{a^{\Phi_\xi(\hat{\eta}_a)}}   \geq    \rho_a \rho_{a^\eta} (h(\omega^\alpha(a^\xi)) \rho_{a^\xi})^{-1}$, we can bound 
\bq
\tau \leq 4 \frac{ \eta^*}{h^*} e^{\beta 2 \overline{\epsilon}} \Sp  \max_{\hat{\xi}}  \sum_{\hat{\eta}_a \in \Gamma(\hat{\xi})}  \frac{1}{2^N }  \rho_{a^{\Phi_\xi(\hat{\eta}_a)}}.
\eq
Furthermore, the map $\Phi_\xi$ is injective for every edge $\hat{\xi}$, thus the sum over all canonical paths passing through edge $\hat{\xi}$ can at most be 
\be
\sum_{\hat{\eta}_b \in \Gamma(\hat{\xi})}  \frac{1}{2^N }  \rho_{b^{\Phi_\xi(\hat{\eta}_b)}} \leq \sum_{\eta \in \bZ_2^{2N}} \frac{1}{2^N} \rho_{b^\eta} .
\ee
Given the Gibbs state $\rho = \sum_a \rho_a P(a)$, we have that the sum 
\be\label{resum}
   \sum_{\eta \in \bZ_2^{2N}} \frac{1}{2^N} \rho_{a^\eta} = 1, 
\ee
evaluates to unity for all syndromes $a$. This follows from the trace identity $\tr{\rho} \1 = 2^{-N}\sum_{\eta \in \bZ_2^{2N}} \sigma(\eta) \rho \sigma(\eta)$.
Since $\tr{\rho} = 1$, we have that 
\be
\1 = \frac{1}{2^N} \sum_{\eta } \sigma(\eta) \rho \sigma(\eta) = \sum_{a,\eta} \frac{1}{2^N} \rho_{a^\eta} P(a). 
\ee
Now since we sum over $\eta$ in the full Pauli group, any error syndrome $a$ can be attained. Hence one can choose some $\tilde{\eta}$ with $E\tilde{\eta} = a$ 
and shift the index accordingly to obtain
\be
	\1 = \left(\sum_\eta \frac{1}{2^N} \rho_{a^\eta} \right) \sum_a P(a).
\ee 
Since we have that $\sum_a P(a) = \1$, we are left with Eqn. (\ref{resum}). Therefore we can always upper bound the maximum by unity and obtain the final bound, so that 
\be
	\tau \leq 4\frac{\eta^*}{h^*}\exp(\beta 2 \overline{\epsilon}).
\ee
Recall that the support number is related to the spectral gap as $\tau = \lambda^{-1}$ which yields the lower bound as stated in the theorem. \vspace{0.5cm}\qed}

The three constants which need to be evaluated for the bound are the maximal path length $\eta^*$, the smallest transition rate $h^*$ and most importantly the generalized energy barrier $\overline{\epsilon}$. The maximal path length will depend on the choice of $\Gamma$ but behaves in general as $\eta^* = \cO(N)$. The smallest transition rate $h^* = \min_{\omega^\alpha(a)} h^\alpha(\omega^\alpha(a))$ is dependent on the bath the system Hamiltonian couples to. In our analysis, however, we only need a few properties of the transition rates to derive the bound. First, we need that these functions satisfy the KMS condition (\ref{KMS}). Second, we need that the lowest rate is a positive, system size independent, constant $h^* > 0$. We will assume that $h^* \sim e^{-\beta\Delta}$, where $\Delta$ denotes the largest eigenvalue difference of $H$ that can be generated by a weight one Pauli, to evaluate the bound. For all local models $\Delta$ is bounded by a constant. The bound on the gap therefore scales like $\lambda \geq \cO(N^{-1}e^{2\beta \overline{\epsilon}})$ in the system size $N$. So if $\overline{\epsilon}$ is a constant in the system size, this lower bound decays as $N^{-1}$. Recall that the mixing time bound we can obtain from the trace norm bound in theorem \ref{Mixingtimebound} already scales as $t_{mix} \leq \cO(N \lambda^{-1})$ the additional factor $N$ does not appear to be significant for the qualitative behavior. Hence, together with theorem \ref{Gen-bound} the final bound on the mixing time is $t_{mix} \leq \cO\left( N^2 e^{2 \beta \overline{\epsilon}}\right)$. Note that in this bound, the only Hamiltonian dependent quantity is the generalized energy barrier $\overline{\epsilon}$. Moreover, we see that this bound is very similar to the phenomenological Arrhenius law $t_{mem} \sim e^{\beta E_B}$. It is therefore the generalized energy barrier that determines whether the systems thermalizes rapidly. That is whether the thermalization time scales as a low degree polynomial in the system size, as opposed to a slow thermalization which is indicated by an exponential  (or possibly a very high degree polynomial) scaling in the system size.\\

\subsubsection{A system size independent lower bound to the spectral gap:} 
It turns out that under particular circumstances, it is possible to remove the $\cO(N^{-1})$ dependence in the lower bound of theorem \ref{no-syn:support_bound} altogether. This leads to a constant lower bound on the spectral gap of the Davies generator, whenever the generalized energy barrier  $\overline{\epsilon}$ is constant. In fact, we are convinced that the $N^{-1}$ prefactor is an artifact of how the lower bound was evaluated for the general bound in theorem \ref{Gen-bound}. Considering the limit $\beta \raw 0$, we observe that the bound (\ref{Gen-bound-eqn}) always  decays as $\lambda \geq \cO(N^{-1})$ for all $H$. However, we know  \cite{kastoryano2014quantum} that the spectral gap  of $\cL_\beta$ has to be constant in the infinite temperature limit for all commuting Hamilonians. The origin of this mild dependence on $N$ becomes clear, when we take a closer look at eqn. (\ref{max-max_bnd}). There, every summand in the original bound was bounded by the largest weight over all edges $\hat{\xi}$. If this step is not performed we obtain a slightly different bound.\\ 

Let us apply the bound $ e^{ \beta 2 \overline{\epsilon}}{h^*}^{-1} \rho_{a^{\Phi_\xi(\hat{\eta}_a)}}   \geq    \rho_a \rho_{a^\eta} (h(\omega^\alpha(a^\xi)) \rho_{a^\xi})^{-1}$, which follows from eqn. (\ref{important-bound}),  directly to eqn. (\ref{Gen-bound-eqn}) in theorem \ref{Gen-bound}. We then have  
\be
\tau \;  \leq \;  \max_{(a,\mu)} \sum_{\hat{\xi} \in \hat{\mu}_a} \frac{4 \; e^{2\beta \overline{\epsilon}}}{ 2^N \; h^*} \sum_{\hat{\eta}_a \in \Gamma(\xi)} \rho_{a^{\phi_\xi(\hat{\eta}_a)}}.
\ee
Recall, that the generalized energy barrier  $\overline{\epsilon}$ was defined as the maximum over all edges, so it is a model dependent constant and can be taken out of the sum. Let us define the parameter
\be\label{def-independent}
	C(\beta) \equiv \max_{(a,\mu)}  \sum_{\hat{\xi} \in \hat{\mu}_a} \frac{1}{ 2^N } \sum_{\hat{\eta}_a \in \Gamma(\xi)} \rho_{a^{\phi_\xi(\hat{\eta}_a)}}.
\ee
If one can show, that this parameter is in fact system size independent, we obtain a lower bound to the spectral gap that only depends on the system size through $\overline{\epsilon}$. For all Hamiltonians $H$, we can always state a lower bound to the spectral gap that is given by 
\be\label{constant-bnd}
	\lambda \geq \frac{h^*}{4 \; C(\beta) } \; e^{-2\beta \overline{\epsilon}}.
\ee
To obtain a constant bound on $\lambda$, one not only needs to check the value of $\overline{\epsilon}$, but also has to evaluate $C(\beta)$. Note, that $C(\beta)$ depends on both, the Hamiltonian though $\rho_a$ and the choice of canonical paths $\Gamma$ due to the sum over $\hat{\eta}_a \in \Gamma(\xi)$. \\

In the limit $\beta \raw 0$  the parameter $C(\beta)$ is easy to evaluate, because we have that $e^{2 \beta \overline{\epsilon}} = 1$ and all $\rho_a = 2^{-N}$. We can therefore choose a set of canonical paths $\Gamma$ that is oblivious to the excitations of the Hamiltonian $H$ and follows a particularly simple protocol: 

Assign a fixed order to all the qubits in the Hamiltonian $H$, and apply for all Pauli operators $\eta$, the single qubit operators $\alpha \in W_1$ following this fixed order.  Any Pauli is build in $t = 1,\ldots,N$ steps, and at the edge $\hat{\xi} = [(a^{\overline{\mu}_{t-1}},\overline{\mu}_{t-1}), (a^{\overline{\mu}_{t}},\overline{\mu}_{t})]$  we have that $t$ - single qubit Pauli's are determined already by $\overline{\mu}_t$.  We can therefore trivially bound $ 2^{-N} \sum_{\hat{\eta}_a \in \Gamma(\xi)} \rho_{a^{\phi_\xi(\hat{\eta}_a)}} \leq 4^{-t}$, since $\rho_{a^{\phi_\xi(\hat{\eta}_a)}} = 2^{-N}$ and we only have at most $4^{N-t}$ undetermined Pauli matrices. Hence $C(0) \leq \sum^N_{t=1}4^{-t} \leq 1/3$ is bounded by a simple geometric series for all $(a,\mu)$ so that eqn.  (\ref{constant-bnd}) reduces to the constant $ \lambda \geq 3/4 \; h^*$. This bound is consistent with constant gap bound obtained in \cite{kastoryano2014quantum} and holds for all commuting Pauli Hamiltonians.\\ 

In light of theorem \ref{Gen-bound}, it is clear that we always have that $0 < C(\beta) \leq \eta^*$, because for every edge we can bound the the summand by unity. If we are, however able to bound every summand by some exponential,  we have that the sum over all links in $C(\beta)$ can be bounded by a constant. In the proof of theorem \ref{Gen-bound} we have shown that $Z = 2^{-N} \sum_{\eta \in \bZ_2^{2N}} \exp(-\beta \epsilon_{a^\eta})$. The evaluation of $C(\beta)$ bears some resemblance to the evaluation of the expectation values in Peierls argument \cite{PSP:2027260}. If we write $C(\beta) = \max_{a,\mu} \sum_{\hat{\xi} \in \hat{\mu}_a} C_{\hat{\xi}}$, and we can estimate in analogy to Peierls argument
\be
	C_{\hat{\xi}} = \frac{1}{2^N}\sum_{\hat{\eta}_a \in \Gamma(\xi)} \rho_{a^{\phi_\xi(\hat{\eta}_a)}} \leq c_0^{|\xi|}e^{- \beta c_1 |\xi| } \Sp \mbox{with} \Sp c_0e^{-\beta c_1} < 1,
\ee
the argument made in the previous paragraph for the limit $\beta \raw 0$ generalizes. This property, however, depends both on the Hamiltonian $H$ and the set of canonical paths $\Gamma$ and one would assume that it has to be checked for every $H$ on a case by case basis. \\

If we are allowed to make some assumptions on the form of the canonical paths, more general results can be proved, and we can improve the bound from theorem \ref{Gen-bound} so that it only depends on $\overline{\epsilon}$. We consider a set of canonical paths $\Gamma_1$, with the property that for all $\hat{\eta}_a \in  \Gamma_1$  \emph{every qubit is addressed only once}, i.e. we find for every $\eta \in \bZ_2^{2N}$ a decomposition $\eta = \oplus_{i \in \Lambda} \alpha_i$ where $\Lambda$ is a labeling of all qubits.

Note that the restriction on $\Gamma_1$ may in many cases lead to a very poor bound on $\overline{\epsilon}$ as can be seen in the toric code example provided in the next section \ref{examples}, where 
Pauli paths are needed that can be of length $2N$ and traverse every single site twice in order to obtain a constant energy barrier.

\begin{theorem}\label{special-bound}
For any commuting Pauli Hamiltonian $H$,  eqn. (\ref{CommPauli}), for which the generalized energy barrier $\overline{\epsilon}$, defined in (\ref{GenEng}), can be evaluated with canonical paths $\Gamma_1$, that address every qubit only once, the spectral gap of the Davies generator is bounded by
\be\label{special-bound-eqn}
	\lambda \geq  \frac{h^*}{4} \exp(- 2 \beta \; \overline{\epsilon} ).
\ee
\end{theorem}

\proof{ The argument is based on a simple extension of he proof given in theorem \ref{Gen-bound}. As was shown in the previous paragraph, we can apply eqn. (\ref{important-bound}),  directly to eqn. (\ref{Gen-bound-eqn}) in theorem \ref{Gen-bound}, so that
\be\label{temp-res}
 \tau \leq C(\beta) \frac{4}{h^*} e^{-2\beta \overline{\epsilon}} \Sp \mbox{with,} \Sp C(\beta) = \max_{(a,\mu)}  \sum_{\hat{\xi} \in \hat{\mu}_a} \frac{1}{ 2^N } \sum_{\hat{\eta}_a \in \Gamma(\xi)} \rho_{a^{\phi_\xi(\hat{\eta}_a)}}.
\ee 
Now, recall that we can write the partial trace over a subset $\cX \subset \mbox{supp}(H)$ as $\ptr{\cX}{A} =  2^{-|\cX|} \sum_{\eta \in \mbox{supp}(\cX)} \sigma(\eta) A \sigma(\eta)$. For any $A$ we naturally have that $\ptr{\cX}{A} = A_{{\cX}^c} \otimes \1_{\cX}$. Furthermore, if $A \geq 0$ is positive semi - definite, we have the inequality $\tr{A}\1 \geq A$ as an operator inequality. Since we 
have that $\tr{A} \1 = \mbox{tr}_{\cX^c} \circ \ptr{\cX}{A}$, we can state the inequality $Z \1 = \tr{\exp(-\beta H)} \1 \geq \ptr{\cX}{\exp(-\beta H)}$.  Both matrices are diagonal in the same basis, given by the projectors $P(a)$, and we have for all eigenvalues labeled by $a$ the inequality
\be
	Z = \frac{1}{2^N} \sum_{\eta \in \bZ_2^{2N}} e^{-\beta e_{a^\eta}} \geq \frac{1}{2^{|\cX|}} \sum_{\eta \in \mbox{supp}(\cX)} e^{-\beta e_{a^\eta}}.
\ee
This implies in particular, that for some chosen subset of qubits $\cX$, we have that 
\be
	\frac{1}{2^N} \sum_{\eta \in \mbox{supp}(\cX)} \rho_{a^\eta} \leq \frac{1}{2^{N-|\cX|}}.
\ee
Now, we consider the canonical paths $\Gamma$ as defined in the theorem, where every qubit is only addressed only once in each canonical path.  We observe, that for every  path $ \hat{\eta}_a \in\Gamma(\xi)$ that uses the edge $\hat{\xi} = [(a^\xi,\xi) , (a^{\xi \oplus \alpha},{\xi \oplus \alpha}) ]$, the single Pauli operator supported on $\mbox{supp}(\xi \oplus \alpha)$ are already determined by the edge. Hence, all paths can only differ on the remaining qubits. We can therefore bound
\be
\frac{1}{ 2^N } \sum_{\hat{\eta}_a \in \Gamma(\xi)} \rho_{a^{\phi_\xi(\hat{\eta}_a)}} \leq \frac{1}{ 2^N } \sum_{\eta \in \mbox{supp}(\xi\oplus\alpha)^c} \rho_{a^\eta} \leq 2^{-|\mbox{supp}(\xi \oplus \alpha)|}.
\ee
We are therefore again left with a geometric series for $C(\beta) = \sum_{t=1}^{N} 2^{-t} \leq 1$. Hence, from this and eqn. (\ref{temp-res}) the inequality (\ref{special-bound-eqn}) follows. \qed}
 
\subsection{Examples}
\label{example}

We consider two simple examples, for which bounds on the spectral gap have been obtained previously \cite{alicki2009thermalization,kastoryano2014quantum}. These examples are chosen to illustrate how the constants in both theorem \ref{Gen-bound} and theorem \ref{special-bound} can be evaluated. 

\subsubsection{The toric code} 
\begin{figure}[h]
\begin{center}
\resizebox{0.85\linewidth}{!}
{\includegraphics{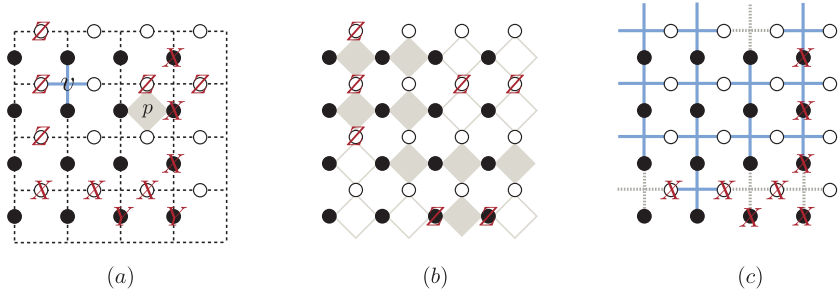}}
\caption{\label{fig:toric_code} (color online) Figure (a) depicts the toric code lattice, where the horizontal (white) qubits as well as the vertical (black) qubits reside on the edges of the lattice. Periodic boundary conditions are assumed. The plaquette term labeled by $p$ are depicted by (grey) rhombi, whereas the vertex operators are depicted by (blue) crosses labeled by $v$. A (red) Pauli operator comprising of local $X,Y,Z$ operators can be decomposed according to their $Z$, c.f. figure (b), and $X$ contribution, figure (c). A Pauli $Y$ contributes to both $X$ and $Z$.}
\end{center}
\end{figure}

To illustrate the bound, let us consider the toric code Hamiltonian \cite{kitaev2003fault} on a square lattice with periodic boundary conditions. For an illustration consider Fig. \ref{fig:toric_code} (a). Every link of the square lattice is fitted with a spin-$1/2$ degree of freedom with Hilbert space $\cH_i = \bC^2$.  The Hamiltonian can be expressed as a sum over plaquettes $\{p\}$ and a sum over the vertices of the lattice $\{v\}$. 
\be\label{toricCode}
	H = -J \sum_p \prod_{i \in p} X_i  - J\sum_v  \prod_{i \in v } Z_i 
\ee
The plaquette terms, marked as (grey) rhombi in Fig. \ref{fig:toric_code} (a), are given as the product of four $X$ operators $\prod_{i \in p} X_i$. Whereas the vertex terms indicated by (blue) crosses are given by the product of $\prod_{i \in v } Z_i$. These multi qubit Pauli operators comprise the generating set $\cG$. \\

Let us now discuss of how to evaluate the bound stated in theorem \ref{Gen-bound}. We first discuss the estimation of $\overline{\epsilon}(\eta)$ as given in eqn. (\ref{energy:bnd}) for an example Pauli operator as shown in (red) letters on the lattice in Fig. \ref{fig:toric_code}(a). It will then become clear that the obtained bound on $\overline{\epsilon}(\eta)$ has to hold in fact for all $\eta$. 

We need to find a suitable set of canonical paths $\Gamma$ to evaluate the bound. As discussed in section \ref{sec:bound}, we only need to construct the Pauli paths that start from identity, since the construction of $\hat{\eta}_a$ is independent of the initial syndrome $a$. This corresponds to choosing a particular order in which single qubit Pauli operators are applied for every $N$-qubit Pauli. Note, that the bound is stated in terms of the optimal choice of paths.  Any other set of paths will also lead to a valid bound. However, this sub optimal choice of $\Gamma$ will naturally yield a looser bound on the spectral gap. 
Given the definition of $\overline{\epsilon}(\eta)$ in (\ref{energy:bnd}), we need to sum the syndromes, i.e. number of generators in $\cG$ that anti commute with $\eta$. This sum is modified by factors $\overline{e}_k(\eta)$, which sets the contributions to zero at which $\eta$ already generated an excitation. As discussed in section \ref{sec:bound} paragraph (a), we therefore need to remove the generators from $\cG$ in the sum that are already violated by $\eta$. This is illustrated in Fig \ref{fig:toric_code} by the removed (highlighted) rhombi in figure (b) and the dashed (grey) vertices in figure (c). The remaining plaquette and vertex terms comprise the reduced generating set $\cG_\eta$. A simple interpretation is now that $\eta$ behaves as a 'logical' operator for the modified code $\cS_\eta$.\\

{\it Constructing the canonical paths $\Gamma$:}  The toric code is a CSS code, which means that the Pauli $X$ and $Z$ contributions occur only in different summands. Since the $X$-type stabilizers only anti commute with the $Z$ contribution of the Pauli $\eta$ and conversely the $Z$-type stabilizers only anti commute with the $X$ contribution, we will split up the $X$ and $Z$ factors in the construction of the paths in $\Gamma$. That is, we write for any $\eta = \eta^x \oplus \eta^z$ and build up $Z$ factors according to Fig.~\ref{fig:toric_code} (b) first before building up any $X$ factors, c.f. Fig.~\ref{fig:toric_code}(c). Naturally we only have to evaluate the $Z$-Paulis on the $X$-stabilizer and vice versa . A local Pauli $Y_i$ is depicted as first applying $X_i$ and then applying $Z_i$ in accordance with $(1,1)_i = (1,0)_i \oplus (0,1)_i$. We thus have to cover the lattice twice, but each subset of stabilizers can be treated independently. Moreover, we can decompose any general Pauli error of either $X$-type or $Z$-type into products one dimensional strings. We will now discuss the form of these strings.

We consider the example Pauli operator in Fig.~\ref{fig:toric_code}. Let us first only focus on the $Z$ - contribution of the Pauli, c.f. Fig \ref{fig:toric_code}(b).  We can first traverse all horizontal lines only using the white qubits and the vertical lines addressing only the black qubits. For the $X$ - contribution in Fig \ref{fig:toric_code}(a)  the role of black and white qubits is reversed. This way plaquette violation are only generated at the end of the string, and these violations vanish, or do not contribute (since $\overline{e}_k(\eta) = 0$), once the string is complete.  We observe, that these strings correspond to minimal error paths of logical Pauli operators \cite{bravyi2009no}, and can be seen as a pair of excitations, that once wrapped around the lattice with periodic boundary conditions constitute a logical operator.\\

In general, one can interpret every violation of the generators  at the end of a string like operator as an excitation of the toric code and all these individual strings of connected $Z$-operators as 'unfinished' logical operators. The general Pauli operator is then a product of trajectories of these individual excitations. It is well known \cite{dennis2002topological}, that excitations in the toric code can be moved without additional energy and the only contribution is at then end of each string. The largest contribution to $\overline{\epsilon}(\eta)$ comes from a string that commutes with the stabilizer group, i.e. is a 
logical operator, because both ends need to be considered. It is therefore at most $2J$. Following this procedure we see that this bound has to hold in fact for any $\eta$, since at every step only a single one dimensional trajectory is build up. Hence, we are left with a bound given by $\overline{\epsilon}(\eta) \leq 2J$. for all $\eta$. 

Now, recall that we have to traverse the lattice twice with this construction, once for the $X$ Contribution and once for the $Z$ Contribution. We therefore have  that the longest canonical paths with our choice of $\Gamma$ has length $\eta^* = 2N$. This leads according to theorem \ref{Gen-bound} to the following lower bound  $\lambda \geq \frac{h^*}{8N} e^{-\beta 4J }$.\\
 
{\bf Remark:} In the construction of the Pauli operators, we have traced the trajectories of individual excitations by applying single qubit Pauli operators. Alternatively we could have generated many excitations by traversing say all plaquette terms horizontally on only a single white qubit before correcting the error in a second run. This way we would have generated a syndrome that would contribute in the order of $2L$ to the energy barrier $\overline{\epsilon}$, since we do not correct for the local excitations which we have created. Here we assume, that the length of the lattice is $L$. This would have lead to an exponentially worse lower bound that would behave as $\lambda \geq \cO\left(L^{-2}e^{-4\beta L}\right)$. This indicates, that a good choice of $\Gamma$ is important. Our choice of canonical paths traverses the lattice twice for Pauli operators that contain $Y$ contributions. Hence, we can not apply theorem \ref{special-bound} to obtain the bound. We therefore stick to the more conservative estimate of theorem \ref{Gen-bound}.

\subsubsection{Commuting Pauli spin chain Hamiltonians}

\begin{figure}[h]
\begin{center}
\resizebox{0.85\linewidth}{!}
{\includegraphics{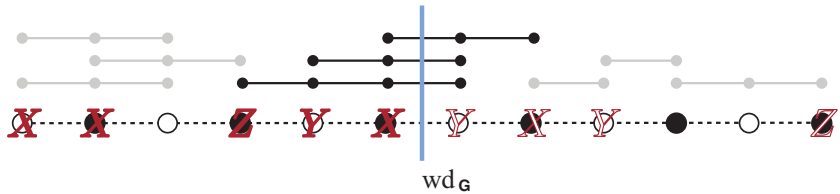}}
\caption{\label{fig:1D_chain} (color online) Sketch of  a one dimensional commuting Pauli Hamiltonian on a line $\Lambda$. The one- to four- local interaction terms are given by the dotted line on top of the one dimensional graph $\Lambda$. The picture shows a Pauli operator $\eta$ in red. The bold Pauli operators correspond to an intermediate Pauli $\overline{\eta}_t$ that violates ${\bf wd}_{\cG} = 3$ interaction terms in the Hamiltonian, whereas the light Pauli operators correspond to the remaining terms that need to be constructed for $\eta$.}
\end{center}
\end{figure}

We consider commuting Pauli Hamiltonains defined on a graph $\Lambda$ that is one-dimensional, i.e. a circle (PBC) or a line (OBC), c.f. Fig. \ref{fig:1D_chain}.  Examples for such system are for instance the one-dimensional Ising model on a length $N$ spin chain given by the Hamiltonian $H_{I} = -J \sum_{i=1}^{N-1} Z_{i} Z_{i+1}$, or the one-dimensional cluster state Hamiltonian ,  also for $N$ spin -$1/2$ particles $H_{C} = -J \sum_{i=2}^{N-1} Z_{i-1}X_iZ_{i+1}$ arranged on a line.  For both the Ising model \cite{alicki2009thermalization} and the cluster State Hamiltonian \cite{temme2014hypercontractivity}, the Davies generator has been derived and lower bounds to the spectral gap were provided. Both bounds are independent of the system size. In Ref \cite{kastoryano2014quantum} a relationship between the constant gap of the Davies generator for all one-dimensional commuting Hamiltonians and the clustering of correlations in the Gibbs state was established.

We now show, that a constant gap lower bound for one-dimensional models can also be derived directly from theorem \ref{special-bound} with a suitable choice of canonical paths $\Gamma$. The choice 
for $\Gamma$ in a one dimensional system is trivial. We choose to decompose every Pauli operator $\eta$ into single qubit Paulis $\alpha \in W_1$ and apply these single site operators in a fixed order, which is the same for every Pauli and canonical for a one-dimensional lattice $\Lambda$ ( we address every qubit once going from left to right ). That is we choose $\eta = \oplus_{i \in \Lambda} \alpha_i$ and construct $\overline{\eta}_t = \oplus_{i=1}^t \alpha_i$. From the Pauli paths $\overline{\eta}$ the canonical paths $\hat{\eta}_a$ follow immediately. This means that for all canonical paths $\hat{\eta}_a \in \Gamma$, every qubit is addressed only once, and the maximal path length is $\eta^* = N$. Hence, we are meeting all the requirements of theorem \ref{special-bound}, so that we can apply the lower bound $ \lambda \geq h^* 4 ^{-1} e^{-2\beta \overline{\epsilon}}$.\\

 Let us define the \emph{width} of the generating set $\cG$ on $\Lambda$ as  
 \be 
 {\bf wd}_\cG = \max_{(i,i+1)  \in \Lambda} \# \{ g \in \cG | (i,i+1) \in \mbox{supp}(g)\},
 \ee 
where the maximum is taken over all edges in the one dimensional interaction graph $\Lambda$. That is we define the width as the largest number of generators that are jointly supported on a single edge of the line, c.f. Fig. \ref{fig:1D_chain}. 

For open boundary conditions we have for any Pauli operator $\eta$ with the previously discussed choice of $\Gamma$ that $\epsilon(\eta) \leq \max_{t \in \Lambda} 2\sum_{k} |J_k| e_{k}(\overline{\eta}_t)\overline{e}_k(\eta) \leq 2J^*{\bf wd}_\cG$, where $J^* = \max_k |J_k|$. For all $i \leq t$ the Pauli's $\overline{\eta}_t$ and $\eta$ coincide, and for all $i > t$, we have that $\overline{\eta}_t$ acts as the identity.  Hence, all $g_k \in \cG$ that are fully supported on either $\{1,\ldots,t\}$ or $\{t+1,\ldots,N\}$, the syndrome compliment $\overline{e}_k(\eta)$ or the syndrome itself  $e_k(\overline{\eta}_t)$ vanishes. Hence only the generators that are supported on both intervals simultaneously can contribute. Hence, these operators need to be supported over the edge $(t,t+1) \in \Lambda$. This number is of course bounded by ${\bf wd}_\cG$. For closed boundary conditions, this scenario occurs at two points along the chain, so that one can see easily that we have that  $\epsilon(\eta) \leq 4J^*{\bf wd}_\cG$. Since these bounds are independent of the Pauli $\eta$ we have an upper bound on $\overline{\epsilon}$. We can state  therefore the following bound for all one-dimensional systems
\bq
	\lambda_{OBC} \geq \frac{h^*}{4} e^{-4 \beta J^*{\bf wd}_\cG} \Sp \mbox{and} \Sp  \lambda_{PBC} \geq \frac{h^*}{4} e^{-8 \beta J^*{\bf wd}_\cG}.
\eq
We obtain the spectral gap for the one-dimensional Ising model from ${\bf wd}_{I} = 1$ and for the cluster state Hamiltonian from ${\bf wd}_{C} = 2$.

\section{Discussion}
We have derived a universal lower bound to the spectral gap of the Davies generator for a commuting Pauli Hamiltonian that is weakly coupled to a thermal heat bath. The bound on the spectral gap establishes a connection between the frequently considered energy barrier for stabilizer codes \cite{bravyi2009no} and the thermalization time of the system. This result can be interpreted as a proof of the phenomenological Arrhenius law, and shows that this law serves in essence as upper bound to the memory time. The bound on the gap as stated in this paper and the naive life time estimate $\tau \sim \lambda^{-1}$ as assumed by the Arrhenius law differ by a factor of $N$, when no further assumptions about the model can be made. To obtain a constant lower bound to $\lambda$, more details about the model are needed. In light of the fact, that the mixing time bound obtained from theorem \ref{Mixingtimebound}, already scales as $t_{mix} \sim \cO(N\lambda^{-1})$, this additional factor $N$ seems insignificant. The crucial conclusion remains unaltered:  {\it Although the existence of an energy barrier is not sufficient to establish thermal stability of the memory,  it is certainly necessary}. 

It is important to point out, that given the spectral gap and the associated mixing time bound we can only make statements about the system's ability to store classical information. This means that the system's ability to reliably store a qubit may have been lost much before thermalization occurs.  The bound is only able to estimate the thermalization time of the system. A good example for this is the toric code in three dimension. A careful analysis of the generalized energy barrier $\overline{\epsilon}$ yields for this model a lower bound to the gap that scales as $\lambda \geq \cO(L^{-3} e^{-4\beta L})$, when the $N$ qubits are arranged on a $N = L\times L \times L$ lattice. We see that this bound predicts a mixing time bond which is exponential in the system size. However,one can see, c.f. Ref.\cite{yoshida2011classification}, that the three dimensional toric code is not a stable quantum memory. The exponential mixing time bound given, however, agrees with the observation that the toric code in three dimension can serve as a stable classical memory. 

For the types of models considered here, one expects a phase transition at some finite $\beta_c$, at which the gap should become independent of the system size. It is an interesting open problem to find a lower bound that is in fact able to reproduce this behavior and indicate a phase transition at some finite temperature. 

The presented approach to bounding the spectral gap of the considered quantum mechanical semi-group is very specific to both the Davies generator and the assumption that the system is described by a stabilizer Hamiltonian. Lower bounds to the spectral gap of more general Davies generators can only be derived under strict assumptions on the Hamiltonians spectrum \cite{temme2013lower}. It is never the less conceivable that the approach presented here can be extended to more complicated systems, such as for instance to the Davies generator of quantum double models \cite{kitaev2003fault} or other semi-groups with an interesting group structure.  

\section{Acknoweldgements}
I would like to thank Anna K\'om\'ar for helpful discussion and for pointing out a mistake in a previous draft, as well as Olivier Landon-Cardinal, Michael J Kastoryano and Fernando Pastawski for helpful discussions. This work was supported by the Institute for Quantum Information and Matter, a NSF Physics Frontiers Center with support of the Gordon and Betty Moore Foundation (Grants No. PHY-0803371 and PHY-1125565). 

%\bibliography{thebib}{}
%\bibliographystyle{unsrt}

\end{document}